\documentclass[prd,aps,11pt,reprint,nofootinbib,superscriptaddress]{revtex4-2}

\usepackage[utf8]{inputenc}
\usepackage[english]{babel}
\usepackage{graphicx}   
\usepackage{dcolumn}    
\usepackage{bm} 
\usepackage{amsmath}    
\usepackage{verbatim}   
\usepackage{color}      
\usepackage{hyperref}
\usepackage{physics}
\usepackage{amssymb}
\usepackage{float}
\usepackage{booktabs}
\usepackage{bigints}
\usepackage{comment}

\newcommand{\ham}{\mathcal{H}}

\begin{document}

\title{Bouncing Bianchi Models with Deformed Commutation Relations}
\date{\today}
\author{Gabriele Barca}\email{gabriele$_$barca001@ehu.eus}
\affiliation{School of Mathematical and Physical Sciences, University of Sheffield, Hicks Building, Hounsfield Road, Sheffield S3 7RH, United Kingdom}\affiliation{Department of Physics, University of the Basque Country, Sarriena Auzoa, 48940 Leioa, Biscay, Spain}
\author{Steffen Gielen}\email{s.c.gielen@sheffield.ac.uk}
\affiliation{School of Mathematical and Physical Sciences, University of Sheffield, Hicks Building, Hounsfield Road, Sheffield S3 7RH, United Kingdom}

\begin{abstract}
We study the anisotropic Bianchi I and Bianchi II models in vacuum in the framework of deformed commutation relations (DCRs). Working in a parametrisation of the spatial metric by a volume and two anisotropy variables, we propose modified Poisson brackets that for the volume alone reproduce the bounce dynamics of effective loop quantum cosmology (LQC), with additional modifications for anisotropy degrees of freedom. We derive effective Friedmann equations and observe cosmological bounces both in Bianchi I and Bianchi II. For Bianchi II, we find that the cosmological bounce now interacts with the usual reflection seen in the Kasner indices in various interesting ways, in close similarity again with what had been seen in LQC. This suggests that the DCR framework could model more general quantum-gravity inspired bounce scenarios in a relatively straightforward way.
\end{abstract}

\maketitle

\section{Introduction}
One of the main problems of classical general relativity is the prediction of spacetime singularities, points in spacetime where curvature scalars diverge and Einstein's field equations cease to be predictive \cite{SingularityTheorems1,SingularityTheorems2}. Significant efforts in modern physics are oriented towards understanding how to remove such singularities, and due to the high-energy regimes reached close to them, it is often expected that this requires a quantum theory of gravity.

The precise mechanism by which singularities might be resolved then depends on the theory and the assumptions made to obtain a manageable model. One particularly popular approach is loop quantum cosmology (LQC) \cite{BojowaldOriginalLQC,LQC2011Review}, in which spatially homogeneous spacetimes are quantised using methods of loop quantum gravity (LQG) \cite{Thiemann_2007,LQGreview}. A generic result in LQC is the replacement of cosmological singularities by a Big Bounce, thereby extending the dynamics beyond what would classically be the endpoint of the evolution. Models of LQC  are not derived directly from LQG, but rather arise from a loop-like quantisation of only a few degrees of freedom obtained after classical symmetry reduction. If the same model is written in a different but classically equivalent form before quantisation, a different quantum theory can be obtained \cite{Assanioussi:2018hee,Liegener:2019zgw}. The connection of such models to the full theory therefore remains unclear (for various attempts to strengthen this connection in LQG and related approaches, see \cite{Dapor:2019mil,Oriti:2016qtz, Gielen:2019kae}). A summary of these and various other criticisms can be found in \cite{LQCproblems1}.

Rather than working in a particular approach to quantum gravity, one can also start with a phenomenological setting in which modifications expected from quantum gravity, such as the introduction of a minimal length scale, are incorporated directly into the Hamiltonian description of various systems \cite{Hossenfelder}. One way in which this can be done is via deformed commutation relations (DCRs), a famous example of which is the generalised uncertainty principle representation in its original formulation by Kempf, Mangano, and Mann (KMM) \cite{KMM}, later expanded by others \cite{Maggiore93,ScardigliGUP}. Through the addition of a higher-order momentum term in the standard commutation relations, the KMM formulation yields a correction to the Heisenberg uncertainty relations that implies an absolute minimal uncertainty on position. The same correction appears in relativistic string theory \cite{ST1,ST2}, where the minimal uncertainty is interpreted as a reminder that it is not possible to probe scales smaller than the size of the string itself. The KMM algebra therefore implements the concept of a minimal length. This deformation was later generalised to other forms (see for example \cite{Maggiore93,Battisti}) that introduce other effects expected such as energy cut-offs. Some forms in particular have been shown to reproduce in a straightforward way the  effective dynamics of other theories and models, such as isotropic LQC, brane cosmology, or the emergent universe scenario \cite{Battisti,IJGMMP,BarcaEU}.

Here we consider the deformation of the anisotropic vacuum Bianchi I and II models described in terms of Misner-like variables. We will deform the isotropic bulk variable, i.e., the comoving volume $v$, separately from the anisotropy variables $\beta_\pm$. In particular, the volume will be deformed with what we call the cut-off algebra, which has been shown to reproduce the bounce of effective LQC \cite{Battisti,IJGMMP}. The anisotropies will be deformed in a similar way, but due to the Misner-like description of the Bianchi models and through the imposition of the Jacobi identities for consistency \cite{Maggiore2021,SebyMatteoSemiclassicalGUP}, they will form a two-dimensional space with naturally arising noncommutativity. In analogy with previous works using similar techniques (e.g., \cite{BarcaEU,SebyMatteoSemiclassicalGUP}), we will be working in a semiclassical limit in which the modified commutators are replaced by effective modified classical Poisson brackets. Such an approach is expected to be justified for semiclassical states that remain peaked around a classical trajectory, where effective classical dynamics provide an excellent approximation to the quantum theory, as is the case for effective LQC \cite{Taveras:2008ke}. One might argue that in the particular application to cosmology, only very semiclassical states are of physical interest. The deformations considered can then still include a variety of expected quantum gravity effects such as spacetime discreteness, noncommutativity, or minimal length effects.

The main results of this paper are the derivation of a modified Friedmann equation similar to isotropic LQC and the replacement of the classical singularity with a bounce also in the anisotropic case (although this had already been hinted at previously \cite{IJGMMP}), the interpretation of such a bounce as a new type of Kasner transition, and the presence of different sequences of Kasner transitions and bounces in the Bianchi II model depending on initial conditions. These last two results in particular were also obtained within the context of LQC \cite{WilsonEwingBianchiLQC}, suggesting that these are more general features of quantum bounce cosmologies. Furthermore, throughout the paper we will give brief insights on the use of the anisotropy variables $\beta_\pm$ as relational time; these variables behave similarly to free massless scalar fields, and are hence globally monotonic. The volume as a function of one of these variables is then a gauge-invariant observable, whose evolution can be studied meaningfully in analogy with LQC, LQG and similar approaches. In this context, we find that the deformation introduced for anisotropy variables acts as a global rescaling of the dynamics, meaning that this kind of quantum correction can also be relevant at late times and low energies.

This manuscript is structured as follows. In Sec. \ref{Sec:ClassicalBianchi} we introduce the classical Bianchi models, showing Bianchi I, Bianchi II, and their solutions in different sets of variables and different time coordinates. In Sec. \ref{Sec:DCR} we introduce the formalism of DCRs, first presenting their general properties and then showing the specific deformations that will be used in later sections. In Sec. \ref{Sec:BounceBI} we implement the chosen deformations on the  Bianchi type I model, showing how cosmological singularities are removed and replaced with a bounce and how this bounce can be interpreted as a new type of Kasner transition. Then in Sec. \ref{Sec:DeformedBII} we go on to deform the Bianchi type II model, showing how different initial conditions give rise to various scenarios with a different order for Kasner reflections or quantum bounces. We conclude the paper with a summary and some final remarks in Sec. \ref{Sec:concl}.

We use natural units $\hslash=c=8\pi G=1$.

\section{Classical Bianchi Models}
\label{Sec:ClassicalBianchi}
In this section we briefly present the main different descriptions used for the classical Bianchi models \cite{MisnerGravitation,PrimordialCosmology}.

The line element for a generic Bianchi model is 
\begin{equation}
    {\rm d}s^2=-\mathcal{N}^2(t)\,{\rm d}t^2+\mathcal{A}_{ij}(t)\omega^i\omega^j\,,
\end{equation}
where $\mathcal{N}$ is the lapse function encoding the freedom to choose a time variable, $\omega^i$ are left-invariant one-forms corresponding to the symmetry group of the model, and $\mathcal{A}_{ij}$ is a symmetric matrix encoding the spatial metric. 

In many Bianchi models (so-called class A models) it is possible to assume a diagonal form for $\mathcal{A}_{ij}$,
\begin{equation}
   \mathcal{A}_{ij}(t) = \begin{pmatrix} a_1^2(t) & 0 & 0 \cr 0 & a_2^2(t) & 0 \cr 0 & 0 & a_3^2(t) \end{pmatrix}\,,
\end{equation}
where $a_i$ are directional scale factors, and this form is consistent with the equations of motion.

From the Lagrangian $\mathcal{L}=\frac{1}{2}\sqrt{-g\,}\,R$ where $R$ is the Ricci scalar, one obtains the generic Hamiltonian for a Bianchi model,
\begin{equation}
   \ham=\mathcal{N}\left(\frac{\sum_ia_i^2p_i^2-\sum_{j\neq k}a_jp_ja_kp_k}{4a_1a_2a_3}\,+a_1a_2a_3U\right),
    \label{classHammetricvariables}
\end{equation}
where $p_i$ are the conjugate momenta to $a_i$ and the potential $U$ is linked to the three-dimensional scalar curvature and only depends on $a_i$. For example, in the case of Bianchi I the potential is zero, while in the case of Bianchi II it depends only on $a_3$. The coordinate volume of spatial slices (which we assume are compact) has also been normalised to unity. The Hamiltonian is constrained to vanish.

The metric variables $(a_1,a_2,a_3)$ are not ideal for the Hamiltonian formulation, as transpires from the complicated Hamiltonian \eqref{classHammetricvariables}. Instead, the Bianchi models are often studied in Misner variables, defined as \cite{MisnerMixmaster}
\begin{align}
\begin{split}
    \alpha &=\frac{1}{3}\,\log(a_1a_2a_3)\,,\\
    \beta_+ &=\frac{1}{6}\,\log(\frac{a_1a_2}{a_3^2})\,,\\
    \beta_- &=\frac{1}{2\sqrt{3\,}\,}\,\log(\frac{a_1}{a_2})\,,
\label{Misnertometric}
\end{split}
\end{align}
or conversely
\begin{align}
\begin{split}
    a_1&=\exp(\alpha+\beta_++\sqrt{3\,}\,\beta_-)\,,\\
    a_2&=\exp(\alpha+\beta_+-\sqrt{3\,}\,\beta_-)\,,\\
    a_3&=\exp(\alpha-2\beta_+)\,.
\label{metrictoMisner}
\end{split}
\end{align}
It is clear how the variable $\alpha$ is related to the isotropic bulk, while the variables $\beta_\pm$ parametrise the anisotropies. With these variables, the Hamiltonian for the Bianchi models becomes
\begin{equation}
    \ham=\mathcal{N}\left(\frac{-\pi_\alpha^2+\pi_+^2+\pi_-^2}{12e^{3\alpha}}\,+e^{3\alpha}\,U\right),
\end{equation}
where $\pi_\alpha$ and $\pi_\pm$ are the momenta conjugate to the corresponding variables. Note how the Misner variables render the kinetic term  diagonal.

Still, even these variables are not useful for the deformations that we want to implement later. Therefore we perform one more change of variables from $\alpha$ to the volume $v=a_1a_2a_3=e^{3\alpha}$. Thanks to the canonical relation $\pi_\alpha=3v\pi_v$, our final Hamiltonian is
\begin{equation}
    \ham=\mathcal{N}\left(-\frac{3}{4}\,\pi_v^2v+\frac{\pi_+^2+\pi_-^2}{12v}+v\,U\right)\approx 0\,.
    \label{finalhamiltonian}
\end{equation}

\subsection{Bianchi I -- the Kasner solution}
\label{classKasner}
Let us start with the Bianchi I model in metric variables. The Hamiltonian is \eqref{classHammetricvariables} with the potential $U$ set to zero. Then the equations of motion in synchronous time $t$, i.e., for $\mathcal{N}=1$, are (using the constraint $\mathcal{H}\approx 0$)
\begin{align}
\begin{split}
    \dot{a}_i&=\frac{a_i}{2a_1a_2a_3}(a_ip_i-a_jp_j-a_kp_k)\,,\\
    \dot{p}_i&=-\frac{p_i}{2a_1a_2a_3}(a_ip_i-a_jp_j-a_kp_k)\,,
\end{split}
\end{align}
with $i\neq j\neq k$ in all equations. It is immediate to see that the quantities $C_i=a_ip_i$ are constants of motion. Then the solutions can be easily found:
\begin{equation}
    a_i(t)\propto|t-t_0|^{k_i}\,,\quad p_i(t)\propto|t-t_0|^{-k_i}\,,
\end{equation}
where
\begin{equation}
    k_i=1-\frac{2C_i}{C_1+C_2+C_3}
    \label{kiCirelation}
\end{equation}
and $t_0$ is an integration constant. The $k_i$ are called Kasner indices and obey the relations \cite{Kasner}
\begin{equation}
    \sum_ik_i=1\,,\qquad\sum_ik_i^2=1\,,
    \label{Kasnerrelations}
\end{equation}
where the first relation comes from the geometric properties of the Bianchi models (and is therefore always valid), while the second one comes from the Hamiltonian constraint and can be modified by the addition of matter or other terms. If ordered, in vacuum the indices will be in the ranges
\begin{equation}
    -\frac{1}{3}\leq\,k_3\leq0,\quad0\leq\,k_2\leq\frac{2}{3},\quad\frac{2}{3}\leq\,k_1\leq1.
\end{equation}
Note how one of the indices has to be negative; therefore the Kasner solution, identified by a set of three constant indices obeying the relations \eqref{Kasnerrelations} and in the expanding phase $t\ge t_0$, describes an empty anisotropic model where lengths contract along one of the spatial directions and expand along the other two, while the total volume $v=a_1a_2a_3\propto |t-t_0|$ grows linearly with time, with a singularity $v\to0$ at $t\to t_0^+$. Conversely, in the contracting branch $t\leq t_0$, the direction related to the negative Kasner index will be expanding, the other two contracting, and the total volume contracts linearly towards the singularity $v\to0$ at $t\to t_0^-$. 

A useful time variable is the harmonic time $\tau$ defined by the harmonic equation $\Box\tau=0$, which translates to the lapse choice $\mathcal{N}=v$. The equations of motion and solutions then greatly simplify:
\begin{align}
\begin{split}
    a_i' &=\frac{1}{2}\,a_i(C_i-C_j-C_k)\,,\\
    p_i' &=-\frac{1}{2}\,p_i(C_i-C_j-C_k)\,,
    \end{split}\\
    a_i(\tau)&\propto\exp(\frac{C_i-C_j-C_k}{2}\,\tau)\,,
\end{align}
where a prime indicates a derivative with respect to $\tau$ and again $i\neq j\neq k$. In harmonic time $\tau$, the solutions are simple exponential functions. From relation \eqref{kiCirelation} we find that
\begin{equation}
    C_i-C_j-C_k=-k_i(C_1+C_2+C_3)\,;
\end{equation}
given the signs of the Kasner indices, it is clear how one exponent will always have a different sign from the other two, and we still have the same solutions with two directions expanding/contracting and the third one contracting/expanding. The volume singularity has been moved to $\tau = -\infty$ (expanding solution) or to $\tau = +\infty$ (contracting solution).

Let us now consider Misner-like variables. Starting from Eq.~\eqref{finalhamiltonian}, we can use Hamilton's equations and the constraint $\ham\approx 0$ to obtain the equations of motion (in synchronous time $t$ with $\mathcal{N}=1$):
\begin{align}
\begin{split}
    \dot{v}&=-\frac{3}{2}\,\pi_vv\,,\quad 
    \dot{\pi}_v=\frac{3}{4}\,\pi_v^2+\frac{\pi_+^2+\pi_-^2}{12v^2}=\frac{3}{2}\,\pi_v^2\,,
    \\\dot{\beta}_\pm&=\frac{\pi_\pm}{6v}\,,\qquad
    \dot{\pi}_\pm=0\,.
    \end{split}
\end{align}
The anisotropy momenta are constants of motion. Furthermore, from the first equation and the constraint $\ham\approx 0$ we can derive the Hubble parameter $H$ and the Friedmann equation:
\begin{equation}
    H^2=\left(\frac{1}{3}\,\frac{\dot{v}}{v}\right)^2=\frac{\pi_v^2}{4}=\frac{\rho_a}{3}\,,\quad\rho_a=\frac{\pi_+^2+\pi_-^2}{12v^2}\,.
    \label{FriedmanneqBIclass}
\end{equation}
We can write the contribution from the anisotropies as an effective energy density $\rho_a$, equivalent to that of two free massless scalar fields in an isotropic universe.

The solutions are then easily found to be
\begin{align}
    v(t)&= v_1|t-t_0|\,,\qquad\pi_v(t)=-\frac{2}{3(t-t_0)}\,,\\
    \beta_\pm&=\overline{\,\beta_\pm}+\frac{\pi_\pm}{6v_1}\,{\rm sgn}(t-t_0)\log\Bigl(v_1|t-t_0|\Bigr)\,,\\\quad\pi_\pm&=\text{const},
\end{align}
where $\overline{\,\beta_\pm}$, $t_0$ and $v_1>0$ are integration constants; the Hamiltonian constraint implies $v_1=\frac{1}{2}\sqrt{\pi_+^2+\pi_-^2\,}$. We again see the linear growth (or contraction) in the volume with $v\to0$ at $t=t_0$. By using the change of variables \eqref{metrictoMisner} to go back to the directional scale factors, we can relate the Kasner indices $k_i$ to the constants $\pi_\pm$:
\begin{align}
\begin{split}
    k_1&=\frac{1}{3}\pm\frac{\pi_++\sqrt{3\,}\,\pi_-}{3\sqrt{\pi_+^2+\pi_-^2\,}}\,,\\
    k_2&=\frac{1}{3}\pm\frac{\pi_+-\sqrt{3\,}\,\pi_-}{3\sqrt{\pi_+^2+\pi_-^2\,}}\,,\\
    k_3&=\frac{1}{3}\pm\frac{-2\pi_+}{3\sqrt{\pi_+^2+\pi_-^2\,}}\,,
\end{split}
\label{KasnerIndicestopi-class}
\end{align}
where $\pm={\rm sgn}(t-t_0)$ denotes whether we are in the expanding or contracting branch.

When using harmonic time $\tau$, once the constraint ${\ham\approx 0}$ has been implemented, the equations of motion and the solutions for the volume and the anisotropies become
\begin{align}
    v'&=-\frac{3}{2}\pi_v v^2=v\,c_1\,,\qquad
    \beta_\pm'=\frac{\pi_\pm}{6}\,,\\
    v(\tau)&\propto e^{c_1\tau}\,,\qquad\beta_\pm(\tau)=\overline{\,\beta_\pm}+\frac{\pi_\pm}{6}\,\tau\,.
    \label{BianchiIclassical}
\end{align}
The anisotropies behave linearly, and the volume is again an exponential with a singularity at either $\tau\to-\infty$ or $\tau\to+\infty$. The constant of motion $c_1=-\frac{3}{2}\pi_v v$ satisfies $4c_1^2=\pi_+^2+\pi_-^2$ (and so $|c_1|=v_1$), but $c_1$ can have either sign corresponding to an expanding or contracting solution. In the $(\beta_+,\beta_-)$ plane, the Bianchi I model is equivalent to a free particle moving on a straight line and at constant speed with respect to $\tau$. In the following we will use the name \emph{particle universe} when referring to the trajectory in the $(\beta_+,\beta_-)$ plane.

\subsection{Bianchi II}
Differently from Bianchi I, the Bianchi type II model has a nonzero potential. It is an exponentially steep wall in the $(\beta_+,\beta_-)$ plane parallel to the $\beta_-$ axis:
\begin{equation}
    U=\frac{\exp(-8\beta_+)}{4v^\frac{2}{3}}\,.
    \label{BIIpotential}
\end{equation}
The equations of motion and the Friedmann equation (with $\mathcal{N}=1$) are then given by
\begin{align}
\begin{split}
    \dot{v}&=-\frac{3}{2}\,\pi_vv\,,\qquad
    \dot{\pi}_v=\frac{3}{2}\,\pi_v^2-\frac{e^{-8\beta_+}}{3v^\frac{2}{3}}\,,\\
    \dot{\beta}_\pm&=\frac{\pi_\pm}{6v}\,,\qquad\quad
    \dot{\pi}_+=2v^\frac{1}{3}e^{-8\beta_+}\,,\\
    \dot{\pi}_-&=0\,,
\end{split}
\end{align}
and
\begin{equation}
    H^2=\frac{\rho_a+U}{3}\,.
\end{equation}
The momentum $\pi_-$ is still a constant of motion, but since the potential depends on $\beta_+$, $\pi_+$ is not. Instead a new constant of motion appears:
\begin{equation}
    \Omega_2=6\pi_vv+\pi_+=\text{const.}
\end{equation}
While there is no analytical solution in synchronous time $t$, it is still possible to analyse the system and derive some dynamical properties.

Given that the potential is so steep, it is only relevant during a brief period of time when the solution is close to it. Its effect is to reflect the particle universe off the wall, thus connecting two Bianchi I Kasner solutions. Using the two constants of motion $\pi_-$ and $\Omega_2$, and relating the momenta to the  velocities $\beta_\pm'\propto\pi_\pm$ in the Bianchi I approximation, it is possible to derive the following reflection law \cite{BKL}:
\begin{equation}
    \sin(\theta_i+\theta_f)=2\bigl(\sin\theta_i-\sin\theta_f\bigr),
    \label{reflectionlaw}
\end{equation}
where the initial angle $\theta_i$ and final angle $\theta_f$ of reflection are measured with respect to the perpendicular to the wall, i.e., with respect to a line parallel to the $\beta_+$ axis (see Fig.~\ref{fig:beta+beta-BII} for an example of this reflection).

This reflection becomes even more evident if we use the harmonic time variable $\tau$, which actually allows us to obtain an analytical solution \cite{PrimordialCosmology}. The equations of motion for the volume $v$, the anisotropies $\beta_\pm$ and the non-constant momentum $\pi_+$ become, respectively,
\begin{align}
\begin{split}
    v'&=-\frac{3}{2}\,\pi_vv^2\,,\\
    \beta_\pm'&=\frac{\pi_\pm}{6}\,,\\
    \pi_+'&=2v^\frac{4}{3}e^{-8\beta_+}\,.
\end{split}
\end{align}
Then the solutions are found by eliminating $\pi_v$ through the Hamiltonian constraint:
\begin{align}
    v(\tau)&=e^{\overline{\,\pi_+}\,\tau}\,\sqrt{\frac{\cosh\Bigl(v_k(\tau-\tau_k)\Bigr)}{v_k}\,}\,,\\
    \beta_+(\tau)&=\frac{\overline{\,\pi_+}}{6}\,\tau+\frac{1}{3}\,\log(\frac{\cosh\Bigl(v_k(\tau-\tau_k)\Bigr)}{v_k}\,)\,,\\
    \pi_+(\tau)&=\overline{\,\pi_+}+2v_k\,\tanh\Bigl(v_k(\tau-\tau_k)\Bigr)\,,
    \label{BIIclasssolution}
\end{align}
where $\tau_k$ is the time when the reflection against the potential wall happens, and $v_k$ is related to the constants $\overline{\,\pi_\pm}$ as
\begin{equation}
    v_k=\sqrt{\overline{\,\pi_+}^2-\frac{\overline{\,\pi_-}^2}{3}\,}\,.
    \label{vkrelation}
\end{equation}
These analytical solutions for Bianchi II were derived in \cite{Taub:1950ez}. They are only defined for real and positive $v_k$.

\begin{figure}
    \centering
    \includegraphics[width=0.9\linewidth]{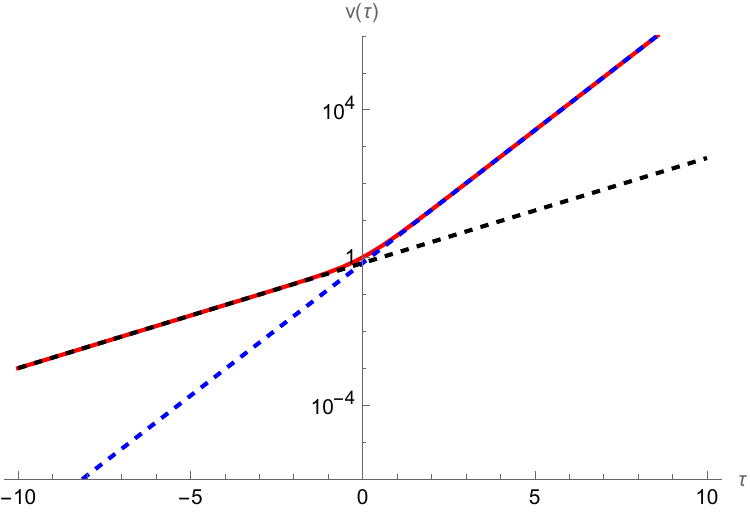}
    \includegraphics[width=0.9\linewidth]{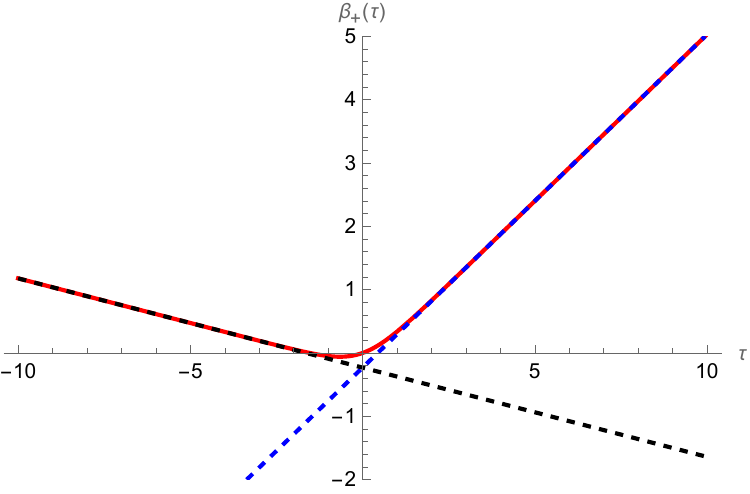}
    \caption{The continuous red lines show the evolution of $v$ (top) and $\beta_+$ (bottom) as functions of $\tau$ in the expanding Bianchi II case, compared with two Bianchi I solutions (shown as dashed blue and black lines).}
    \label{fig:BianchiIIclass}
\end{figure}

\begin{figure}
    \centering
    \includegraphics[width=0.9\linewidth]{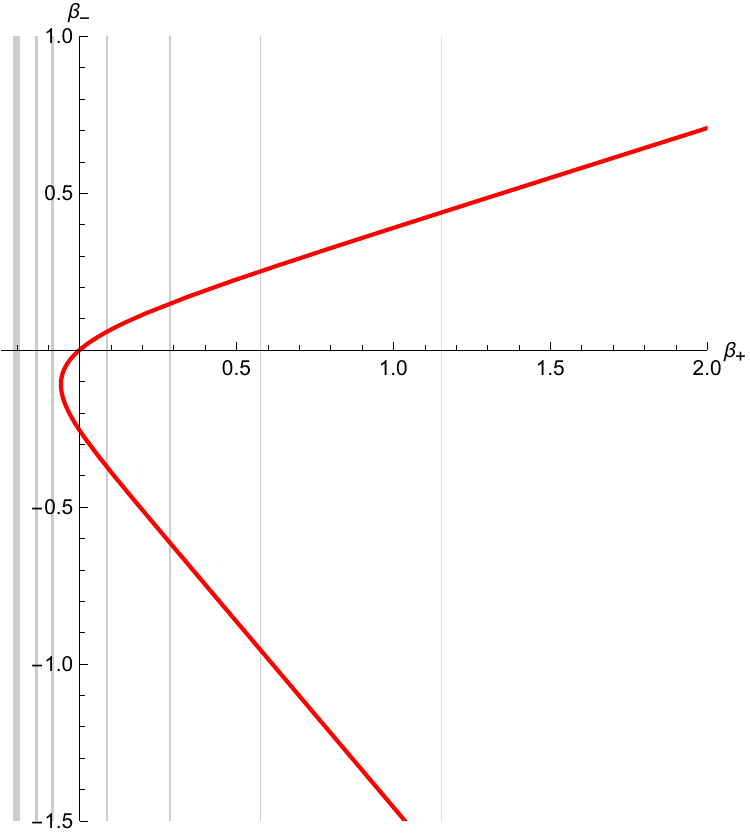}
    \caption{Trajectory of the classical Bianchi II model in the $(\beta_+,\beta_-)$ plane. The vertical grey lines are lines of constant potential, growing towards the left for smaller values of $\beta_+$.}
    \label{fig:beta+beta-BII}
\end{figure}

Because of the term $e^{\overline{\,\pi_+}}$ in the solution for the volume and the fact that $\abs{\overline{\,\pi_+}}\geq v_k$, the sign of $\overline{\,\pi_+}$ dictates whether the solution is expanding or contracting. Figs.~\ref{fig:BianchiIIclass} and \ref{fig:beta+beta-BII} show the dynamics of the expanding Bianchi II solution with respect to harmonic time $\tau$ and the trajectory in the $(\beta_+,\beta_-)$ plane respectively. It is evident how the solutions consist of a smooth transition between two Kasner solutions. It is also possible to see that the potential wall prevents $\beta_+$ from becoming too negative.
When using metric variables, i.e., the three scale factors $a_1$, $a_2$, and $a_3$, the reflection law \eqref{reflectionlaw} becomes a map for the Kasner indices \cite{BKL}: the indices after the reflection (indicated with primes) as functions of the indices before the reflection (without primes) are
\begin{equation}
\begin{aligned}
    &k_1'=\frac{k_1+2k_3}{1+2k_3}\,,\\
    &k_2'=\frac{k_2+2k_3}{1+2k_3}\,,\\
    &k_3'=-\frac{k_3}{1+2k_3}\,.
    \label{classKasnermap}
\end{aligned}
\end{equation}
It is clear that the index $k_3$ and one of the other two change sign. Explicitly, 
\begin{equation}
k_3=\frac{v_k}{2\overline{\,\pi_+}-v_k}\,,\quad k'_3=-\frac{v_k}{2\overline{\,\pi_+}+v_k}\,;
\label{k3pirelation}
\end{equation}
hence (both for an overall expanding and for an overall contracting solution) the effect of the reflection is that the direction $a_3$ goes from expanding to contracting, whereas one of the other two directions that were previously contracting is now expanding. In particular, in the expanding case with $\overline{\,\pi_+}>0$ we have $k_3>0$ and $k_3'<0$, while in the contracting case with $\overline{\,\pi_+}<0$ we have the opposite.

For later reference, it is useful to introduce the quantity $r$ corresponding the ratio between the two momenta, 
\begin{equation}
    r(\tau)=\frac{\pi_+}{\pi_-}=\overline{r}+2\sqrt{\overline{r}^2-\frac{1}{3}\,}\tanh(\overline{\,\pi_-}\sqrt{\overline{r}^2-\frac{1}{3}\,}\,(\tau-\tau_k)),
    \label{ratio}
\end{equation}
where $\overline{r}=\overline{\pi_+}/\overline{\pi_-}$. The ratio $r$ corresponds to the cotangent of the angle between the trajectory in the $(\beta_+,\beta_-)$ plane and the positive branch of the $\beta_+$ axis.

If we want to highlight the transition, we can define the asymptotic values of $r$ as
\begin{equation}
\begin{aligned}
    &r_-=r(\tau\to-\infty)=\overline{r}- 2\, {\rm sgn}(\overline{\,\pi_-})\sqrt{\overline{r}^2-\frac{1}{3}}\,,\\
    &r_+=r(\tau\to+\infty)=\overline{r}+ 2\, {\rm sgn}(\overline{\,\pi_-})\sqrt{\overline{r}^2-\frac{1}{3}}\,.
\end{aligned}
\end{equation}
Then the relation \eqref{vkrelation}, when expressed in terms of $r$, corresponds to the following: for the expanding case,
\begin{equation}
    r_-<\frac{1}{\sqrt{3\,}}\;(\overline{\,\pi_-}>0)\qq{or}r_->-\frac{1}{\sqrt{3\,}}\;(\overline{\,\pi_-}<0)\,,
    \label{r-exp}
\end{equation}
while for the contracting case
\begin{equation}
    r_-<-\frac{1}{\sqrt{3\,}}\;(\overline{\,\pi_-}>0)\qq{or}r_->\frac{1}{\sqrt{3\,}}\;(\overline{\,\pi_-}<0)\,.
    \label{r-contr}
\end{equation}
These conditions ensure that the reflection off the potential wall actually takes place. Indeed, it is possible to see that the relevant equipotential line $\beta_\text{wall}$, calculated as the value of $\beta_+$ that makes the anisotropic density $\rho_a$ defined in \eqref{FriedmanneqBIclass} and the potential $U$ in \eqref{BIIpotential} comparable, grows monotonically with the volume as $\beta_\text{wall}\propto\log(v^\frac{4}{3})$. This means that the wall is moving right (towards greater values of $\beta_+$) in an expanding universe, and left in a contracting one. Then the condition \eqref{r-exp} in the expanding case ensures that the particle universe in the $(\beta_+,\beta_-)$ plane is not moving too fast towards the right to outrun the wall, while the condition \eqref{r-contr} in the contracting case is requiring that the particle universe is moving fast enough towards the left to hit the receding wall.

Finally, we can compute how the transition changes the value of the ratio, i.e., express the reflection map in terms of the asymptotic values of $r$. From ${r_+=r_-+4\,{\rm sgn}(\overline{\,\pi_-})\sqrt{\overline{r}^2-1/3\,}\,}$, we obtain
\begin{equation}
    r_+=\frac{4}{3}\,{\rm sgn}(\overline{r})\sqrt{r_-^2+1\,}-\frac{5}{3}\,r_-\,.
    \label{rtransitionBIIclass}
\end{equation}

These classical dynamics of the Bianchi I and Bianchi II models will be used as comparison for their deformed counterparts.

\section{Deformed Commutation Relations}
\label{Sec:DCR}
As mentioned in the introduction, DCRs are a simple framework to implement on any Hamiltonian system effects expected from more fundamental theories, such as a minimal length or energy cut-offs \cite{Hossenfelder}. We will start by presenting the general properties of the DCRs, and then focus on the specific form that will be used in our study.

\subsection{General properties}

In various examples, it is known that quantum gravitational effects can be introduced through the replacement of trivial commutators with functions of a momentum operator \cite{IJGMMP}
\begin{equation}
    \comm{\hat{q}}{\hat{p}}=i\,f(\hat{p})\,,
    \label{dcreq}
\end{equation}
where $\hat{q}$ and $\hat{p}$ represent two generic conjugate quantum operators. The requirement of satisfying Eq.~\eqref{dcreq} then means that the action of the fundamental operators is modified. It is easier to work in the momentum polarisation with wavefunctions ${\psi=\psi(p)}$; then there are two main possible operatorial representations, depending on which operator is modified: we can define
\begin{equation}
    \hat{q}\,\psi(p)=i\,f(p)\,\psi'(p)\,,\quad\hat{p}\,\psi(p)=p\,\psi(p)\,,
    \label{modqrepresentation}
\end{equation}
or
\begin{equation}
    \hat{q}\,\psi(p)=i\,\psi'(p)\,,\quad\hat{p}\,\psi(p)=g(p)\,\psi(p)\,,
    \label{modprepresentation}
\end{equation}
where the function $g$ is related to $f$ through ${g^{-1}=\int {\rm d}p/f(p)}$. The equivalence between these two operatorial representations is still unclear and, although proven for specific cases \cite{IJGMMP}, seems to fail in general. In any case, a modified Schr\"odinger equation and an overall deformed dynamics are obtained. Furthermore, there might be additional structures (such as a minimal length as mentioned above) that require additional care.

The structure of DCRs becomes more interesting when extended to higher spatial dimensions. In this case the main commutator is generalised to
\begin{equation}
    \comm{\hat{q}_i}{\hat{p}_j}=i\,\delta_{ij}\,f(\hat{p}_\text{tot}),
    \label{higherDcommutator}
\end{equation}
where the function $f$ must depend on the total momentum ${p_\text{tot}^2=\sum_ip_i^2}$ if we want to preserve rotational invariance \cite{Maggiore2021}. However, the Jacobi identities imply that the commutator between different space directions cannot be zero, but must instead be
\begin{equation}
    \comm{\hat{q}_i}{\hat{q}_j}=i\,\frac{f'(\hat{p}_\text{tot})\,\hat{J}_{ij}}{\hat{p}_\text{tot}}\,,
\end{equation}
where an angular momentum operator ${\hat{J}_{ij}=\hat{q}_i\hat{p}_j-\hat{q}_j\hat{p}_i}$ appears. Therefore DCRs also introduce noncommutativity without requiring one to assume it independently. However, note that with this construction it is not possible to define the function $g$ of the representation \eqref{modprepresentation} in a consistent way, and therefore only the representation \eqref{modqrepresentation} where the position operator is modified is available (the derivative is taken with respect to the corresponding momentum component). Furthermore, since the commutator \eqref{higherDcommutator} itself does not commute with the position operators any more, the validity of some of the properties that are commonly used in quantum mechanics is restricted. 

For the purpose of this study we are going to restrict ourselves to the (semi-)classical limit of DCRs, where the modified commutators become modified Poisson brackets, and there are no issues with operator ordering. As we have discussed in the introduction, such an effectively classical description is expected to capture the properties of very semiclassical states that could describe an effectively classical universe in such modified theories. This limit can be defined consistently as a modified symplectic structure \cite{SebyMatteoSemiclassicalGUP}, and has already been used in various contexts in the past. The full quantum analysis remains as a future project, but a first try with a different deformation affecting only the anisotropies has already been attempted \cite{SebyGUP}.

\subsection{Cut-off algebra and Misner variables}

As hinted above, we will use a deformed Poisson algebra which implements a momentum cut-off for $\pi_v$. When implemented in isotropic cosmology, such an algebra can reproduce dynamics similar to LQC and in particular replace the classical singularities with a Big Bounce \cite{Battisti,IJGMMP}.

Following this past work, for the volume we write 
\begin{equation}
    \pb{v}{\pi_v}=b(\pi_v)=\sqrt{1-\mu_v^2\pi_v^2\,}\,.
    \label{LoopAlgebra}
\end{equation}
This choice of $b(\pi_v)$ is singled out by the requirement that in the isotropic setting it reproduces exactly the modified Friedmann equation of effective LQC. Using the Hamiltonian constraint, it is possible to work backwards and derive the required deformation from the modified Friedmann equation, and the result is unique (up to a global sign, which is fixed by consistency in the limit $\mu_v\to0$). This agreement is also evidenced by the fact that in the quantum setting, the eigenvalue $g(p)$ in Eq.~\eqref{modprepresentation} for this deformation is a trigonometric sine function, which would make the modified momentum operator act in the same way as in LQC \cite{LQC2011Review}. Other functions have been studied that are also able to remove singularities, but they result in qualitatively different dynamics such as, e.g., an asymptotic approach to an eternal Einstein-static phase \cite{IJGMMP,BarcaEU}.

Regarding the anisotropy variables, we assume
\begin{equation}
    \pb{\beta_\pm}{\pi_\pm}=\delta_\pm\,f(\pi_\text{tot})\,,\quad\pi_\text{tot}^2=\pi_+^2+\pi_-^2\,,
    \label{DeformedAnisotropies}
\end{equation}
where in analogy with \eqref{higherDcommutator} we impose that the deformation only depends on the total momentum $\pi_\text{tot}$, and
\begin{equation}
    \pb{\beta_+}{\beta_-}=\frac{f'(\pi_\text{tot})\Bigl(\beta_+\pi_--\beta_-\pi_+\Bigr)}{\pi_\text{tot}}
\end{equation}
from self-consistency using the Jacobi identity. In this case, we will keep the function $f$ as generic as possible apart from the assumptions that the deformation only depends on momenta, and that the rotational invariance (here rotations in the $(\beta_+,\beta_-)$ plane) of the classical Bianchi I model is maintained. When necessary for numerical calculations, we will assume that $f$ is of the same form as the function $b$ used for the volume.

\section{Bouncing Bianchi I}
\label{Sec:BounceBI}
Let us start by deforming the simple Bianchi I model. Since the potential is zero and the momenta $\pi_\pm$ are constants of motion, it is clear how the deformation function $f$ in \eqref{DeformedAnisotropies} is also a constant of motion, and will therefore be just a constant rescaling.

Let us start by setting $\mathcal{N}=1$ and using synchronous time $t$. The equations of motion and the Friedmann equation get modified:
\begin{align}
\begin{split}
    \dot{v}&=-\frac{3}{2}\,\pi_vv\sqrt{1-\mu_v^2\pi_v^2\,}\,,\\
    \dot{\pi}_v&=\frac{3}{2}\,\pi_v^2\sqrt{1-\mu_v^2\pi_v^2\,}\,,\\
    \dot{\beta}_\pm&=\frac{\pi_\pm\,f(\pi_\text{tot})}{6v}\,,\,\qquad\dot{\pi}_\pm=0\,,
\end{split}
\end{align}
and
\begin{equation}
    H^2=\frac{\rho_a}{3}\left(1-\frac{\rho_a}{\rho_\mu}\right)\,,
\end{equation}
where $\rho_\mu=3/4\mu_v^2$ is a critical density. This modified Friedmann equation is of a very similar form to effective Friedmann equations that have been derived for semiclassical states in isotropic LQC with a free massless scalar field \cite{Taveras:2008ke}, except that the matter density of the scalar field is replaced by the effective energy density $\rho_a$ associated to anisotropies.\footnote{For clarity we should point out that no similar effective Friedmann equation exists for Bianchi I models in LQC \cite{noFriedmannLQC}.} The correction factor implies a critical point in the dynamics where $\dot{v}=0$ at $\rho_a=\rho_\mu$, and it is able to remove the singularity. Indeed, the solution for the volume is
\begin{equation}
    v(t)=v_1\sqrt{(t-t_0)^2+\frac{v_B^2}{v_1^2}\,}\,,\qquad v_B=\frac{2}{3}\,v_1\mu_v\,.
\end{equation}
The value $v(t_0)=v_B$ constitutes an absolute minimum, and the singularity is replaced by a bounce in the same way as in isotropic LQC. Furthermore, it is trivial to see how in the limit $\mu_v\to0$, $\rho_\mu\to\infty$ and $v_B\to0$, and the classical evolution is recovered.

The solution for the anisotropies is
\begin{equation}
    \beta_\pm(t)=\overline{\,\beta_\pm}+\frac{\pi_\pm\,f(\pi_\text{tot})}{6v_1}\,\log(\frac{v_1}{2}\,(t-t_0)+\frac{v(t)}{2}).
\end{equation}
As mentioned earlier, besides the modification to the argument of the logarithm, the function $f$ acts as a constant rescaling. This will have interesting consequences later.

\subsection{The bounce as a Kasner transition}
An interesting description of the bouncing Bianchi I model can be extrapolated from the dynamics in terms of harmonic time $\tau$.

\begin{figure}
    \centering
    \includegraphics[width=0.9\linewidth]{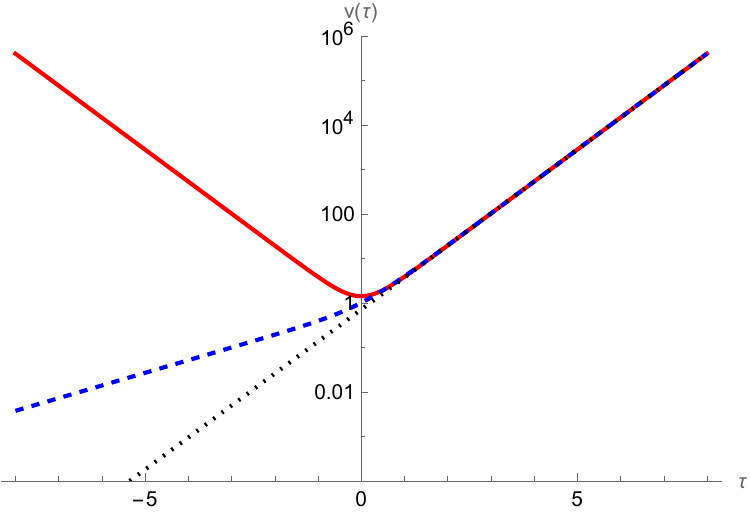}
    \caption{The volume $v$ as function of harmonic time $\tau$ for the bouncing Bianchi I model (red continuous line), compared with the classical Bianchi II (blue dashed line) and Bianchi I (black dotted line) models. Here we have set $\overline{\,\pi_-}=1$, $\overline{\,\pi_+}=2/\sqrt{3\,}$ and $\mu_v=3\sqrt{6/7\,}$; the large value of the deformation parameter was chosen to have the three expanding branches for $\tau>0$ match, highlighting the different smooth transitions.}
    \label{fig:BianchiIBounce}
\end{figure}

The equations of motion for the volume and anisotropies become
\begin{equation}
    v'=-\frac{3}{2}\,\pi_vv^2\sqrt{1-\mu_v^2\pi_v^2\,}\,,\qquad \beta'_\pm=\frac{\pi_\pm\,f(\pi_\text{tot})}{6}.
\end{equation}
The solutions are then easily found as
\begin{equation}
    v(\tau)=v_B\cosh(v_1\tau),\qquad\beta_\pm(\tau)=\overline{\,\beta_\pm}+\frac{\pi_\pm\,f(\pi_\text{tot})}{6}\,\tau.
\end{equation}
Once again, the volume reaches the minimum $v_B$ at one particular value of $\tau$ (which we have chosen as $\tau=0$), and the singularity is removed. This bouncing Bianchi I model can be interpreted as a smooth transition between two classical Bianchi I solutions (\ref{BianchiIclassical}), one contracting and one expanding, as evident from Fig. \ref{fig:BianchiIBounce}. Using the solution for the volume, the Hamiltonian constraint and the relations between $\pi_\pm$, $v_1$ and $v_B$, we can also express the cut-off algebra function $b$ in \eqref{LoopAlgebra} in terms of time $\tau$:
\begin{equation}
   b(\tau)=\sqrt{1-\frac{\mu_v^2(\pi_+^2+\pi_-^2)}{9v^2}}=\sqrt{1-\frac{v_B^2}{v^2}}=\tanh(v_1|\tau|)\,.
\end{equation}
The function goes to zero at the bounce and approaches the classical value $b=1$ far away from it. We can also see that the anisotropies are still linear, but the slope is modified by the presence of the function $f$.

To understand the behaviour of the Kasner indices around the bounce, we can use a time-dependent notion of quasi-Kasner indices as introduced in \cite{deCesare},
\begin{equation}
    k_i(\tau)=\frac{a'_i}{a_i}\,\frac{v}{v'}
\end{equation}
(note that the definition is independent of the choice of time coordinate), where we can recover the directional scale factors through \eqref{metrictoMisner}. We obtain
\begin{equation}
    \begin{aligned}
        k_1(\tau) & = \frac{1}{3} + \frac{(\pi_++\sqrt{3\,}\,\pi_-)f}{3b(\tau)\sqrt{\pi_+^2+\pi_-^2\,}}\,{\rm sgn}\,\tau\,,\\
        k_2(\tau) & = \frac{1}{3} + \frac{(\pi_+ - \sqrt{3\,}\,\pi_-)f}{3b(\tau)\sqrt{\pi_+^2+\pi_-^2\,}}\,{\rm sgn}\,\tau\,,\\
        k_3(\tau) & = \frac{1}{3} - \frac{2\,\pi_+f}{3b(\tau)\sqrt{\pi_+^2+\pi_-^2\,}}\,{\rm sgn}\,\tau\,,
    \end{aligned}
    \label{pitoKasnerindicesdeformed}
\end{equation}
where the time dependence is apparent. Comparing asymptotic limits (where the volume behaves as simple exponentials), we can find a relation between the (asymptotically constant) Kasner indices before and after the bounce, similarly to what we did for the classical Bianchi II case:
\begin{equation}
    k_i'=\frac{2}{3}-k_i,
    \label{BounceKasnermap}
\end{equation}
where as before the primes indicate the indices after the bounce. The same map was found by Wilson-Ewing in \cite{WilsonEwingBianchiLQC} in the context of LQC. We also see that, while the quasi-Kasner indices still sum to one by definition, the second Kasner relation is modified compared to the standard Bianchi I relation,
\begin{equation}
    \sum_ik_i(\tau)^2=\frac{1}{3}+\frac{2}{3}\,\frac{f^2}{b(\tau)^2}\,.
    \label{deformedKasner2}
\end{equation}
In particular this sum, just as the individual $k_i$ themselves, actually diverges as we approach the bounce; also, since $f\neq 1$ in general, the relation is modified nontrivially even in the asymptotic limit.

Changing the second Kasner relation changes the allowed ranges of the indices. However, since $b\approx 1$ everywhere except close to the bounce, in later sections we will only consider the effects of the function $f$. For small values of $f$ it is possible for all three indices to have the same (positive) sign. On the other hand, for large values of $f$, the indices can be much bigger than $1$ in absolute value, but at least one of them will still be negative. This will have consequences when we consider the deformation of the Bianchi II model, but the fact that the indices could be all positive and of similar magnitude might also lead to a possible isotropization effect: notice that in the limit $f\rightarrow 0$ we have $k_i\rightarrow \frac{1}{3}$, a strictly isotropic geometry.

Even though the algebra \eqref{LoopAlgebra} is related to isotropic LQC in the sense that it gives a similar Friedmann equation in the isotropic case, our treatment of anisotropies is different and more general, given that we did not need to specify the function $f$. In the case of Bianchi I, $f$ acts as a constant rescaling of the velocities $\beta'_\pm$ relative to the conserved momenta $\pi_\pm$, which means that Kasner-like solutions are still obtained as soon as $b\approx 1$ away from the bounce, just with potentially different values of $k_i$. The interpretation of a Bianchi I bounce as a new type of Kasner transition is hence a generic property of this kind of scenarios, with (\ref{BounceKasnermap}) a similarly general result.

\subsection{The role of deformed anisotropies}
We have seen how, in the case of Bianchi I, the function $f$ deforming the anisotropies is a constant and acts as a simple multiplicative rescaling for the velocities $\dot\beta_\pm$. It also does not appear at all in the expression $v(\tau)$ for the volume as a function of harmonic time, and so appears to have no impact on the bouncing dynamics. However, the relation $v(\tau)$ is coordinate dependent and does not constitute an observable. It is more meaningful to look at the behaviour of relational observables which express relations between dynamical degrees of freedom \cite{Rovelli:1990ph,relationalobs}.

In isotropic quantum cosmology it is common to use a free massless scalar field as an internal time variable \cite{PrimordialCosmology}. In our model there is no matter, but as mentioned above the two anisotropy variables in the Bianchi I model effectively behave as free scalar fields. It is therefore natural to choose one of them as internal time. This amounts to choosing $\mathcal{N}=6v/\pi_\pm$ in the classical case and $\mathcal{N}=6v/(\pi_\pm f)$ in the deformed case.

In the classical case, the equations of motion for the volume and the Friedmann equation become
\begin{equation}
    \dv{v}{\beta_\pm}=-\frac{9\pi_vv^2}{\pi_\pm}
\end{equation}
and
\begin{equation}
    \left(\frac{1}{3v}\,\dv{v}{\beta_\pm}\right)^2=\frac{\pi_\text{tot}^2}{\pi_\pm^2}\,,
\end{equation}
where as usual the Hamiltonian constraint has been used to eliminate $\pi_v$ and obtain this last equation. As expected, the right-hand side is a constant and the solution for $v(\beta_\pm)$ will be an exponential, similarly to what happens in isotropic cosmology when using a scalar field.

However, when considering the deformed case, the modified Friedmann equation turns out to be
\begin{equation}
    \left(\frac{1}{3v}\,\dv{v}{\beta_\pm}\right)^2=\frac{\pi_\text{tot}^2}{\pi_\pm^2f^2(\pi_\text{tot})}\left(1-\frac{\rho_a}{\rho_\mu}\right)\,.
    \label{relationalFriedmann}
\end{equation}
Here we can see both the correction coming from the deformation of the volume, i.e., the factor $1-\rho_a/\rho_\mu$ that was present also when using other time variables, and the correction due to the deformation of the anisotropies, encoded by the presence of $f^2$ at the denominator. Interestingly, while the former type of deformation is only relevant close to the bounce, the fact that $f$ is a constant of motion means that it affects the dynamics across the entire evolution. This could be seen as a curvature-independent effect coming from quantum corrections, which we will encounter again in the Bianchi II case.

\section{Deformed Bianchi II}
\label{Sec:DeformedBII}
We will now move on to the deformation of the Bianchi II model. While it is impossible to derive full analytical solutions, we can approximate the Bianchi II solution using the general properties we derived for the undeformed Bianchi II and deformed Bianchi I cases studied previously. We will also use numerical integration to visualise particular cases and obtain quantitative results related to the deformations we have introduced.

\subsection{Initial conditions and different scenarios}
From the deformed equations of motion in synchronous time, we obtain the modified Friedmann equation
\begin{equation}
    H^2=\left(\frac{\dot{v}}{3v}\right)^2=\frac{\rho_a+U}{3}\left(1-\frac{\rho_a+U}{\rho_\mu}\right)\,.
    \label{DefBianchiIIFried}
\end{equation}
We can see how the potential $U$ of Bianchi II (introduced in (\ref{BIIpotential})) gets added to the total energy density also in the correction factor.

Now, we have mentioned how in Bianchi II the potential is negligible for most of the dynamics except when $\beta_+$ is close to $\beta_\text{wall}$, i.e., when the anisotropy density $\rho_a$ is comparable to the potential $U$. Similarly, the quantum corrections responsible for the bounce are relevant only when the total energy density is close to the critical value $\rho_\mu$. Furthermore, it is safe to assume that these two situations are not happening at the same time (except for very specific initial conditions). Therefore we can expect the dynamics of the bouncing Bianchi II model to consist of a series of Kasner solutions linked either by a reflection off the wall or by a quantum bounce.

The order in which the different types of transition happen will depend on the initial conditions. In particular, considering the Kasner indices and the two maps \eqref{classKasnermap} and \eqref{BounceKasnermap}, in the case $f=1$ there are three possibilities. Remembering that in a bounce scenario the universe is initially globally contracting, the following can happen (the superscripts here indicate the different values at different times in the evolution):
\begin{itemize}
    \item if the initial Kasner index $k_3^{(0)}$ is positive, then no Kasner reflection happens before the quantum bounce since the scale factor $a_3$ is contracting; then after the bounce, when the universe is now expanding, there will be a single Kasner reflection as in a classical Bianchi II model;
    \item if the initial value of $k_3$ is $-1/3<k_3^{(0)}<-2/7$, a Kasner reflection happens before the bounce since $a_3$ is expanding; then after the reflection $k_3$  becomes ${2/3<k_3^{(1)}<1}$, after the bounce ${-1/3<k_3^{(2)}<0}$, and no other transition happens;
    \item if $-2/7<k_3^{(0)}<0$, a Kasner reflection happens before the bounce, then after the reflection ${0<k_3^{(1)}<2/3}$, after the bounce $0<k_3^{(2)}<2/3$, and there will be an additional Kasner reflection after the bounce bringing the exponent back to ${-1/3<k_3^{(3)}<0}$.
\end{itemize}
These results were also found in \cite{WilsonEwingBianchiLQC}; therefore, as in the discussion above, these might be generic properties of bouncing cosmologies.

In the case of undeformed anisotropies with ${f(\pi_\text{tot})=1}$, and given that we assume we are far enough from the bounce so we can take $b=1$, the relationship between $k_3$ and the ratio $r=\pi_+/\pi_-$ is the same as in classical Bianchi II, see in particular (\ref{k3pirelation}). Hence the three cases can be translated into possible initial conditions for $r$. Remembering that the universe is initially contracting, the potential wall goes towards the left, and the bounce does not change the value of $r$, we have:
\begin{itemize}
    \item if initially $r^{(0)}>-1/\sqrt{3\,}$ (for $\overline{\,\pi_-}>0$) or ${r^{(0)}<1/\sqrt{3\,}}$ (for $\overline{\,\pi_-}<0$), the universe is not moving fast enough towards the left to catch up to the wall, and therefore there will be no Kasner reflection before the bounce; however, after the bounce, when the wall is moving towards the right, it is able to reach the particle universe and according to \eqref{rtransitionBIIclass} (with the plus sign for the expanding case) there will be a single Kasner transition to $r^{(1)}>1/\sqrt{3\,}$ or $r^{(1)}<-1/\sqrt{3\,}$ (for positive or negative $\overline{\,\pi_-}$ respectively), the same as in the expanding case of classical Bianchi II;
    \item if initially $r^{(0)}<-13/(3\sqrt{3})$ (for $\overline{\,\pi_-}>0$) or $r^{(0)}>13/(3\sqrt{3})$ (for $\overline{\,\pi_-}<0$), the universe is able to reach the wall and a Kasner transition happens before the bounce; then, according to relation \eqref{rtransitionBIIclass}, we will have $r^{(1)}>1/\sqrt{3\,}$ or $r^{(1)}<-1/\sqrt{3\,}$ (for positive or negative $\overline{\,\pi_-}$ respectively) and the universe will continue expanding to infinity with a standard Kasner solution;
    \item if initially $-13/(3\sqrt{3})<r^{(0)}<-1/\sqrt{3\,}$ (for ${\overline{\,\pi_-}>0}$) or $1/\sqrt{3\,}<r^{(0)}<13/(3\sqrt{3})$ (for $\overline{\,\pi_-}<0$), the universe is able also here to reach the wall and to undergo a Kasner reflection before the bounce, but then $-1/\sqrt{3\,}<r^{(1)}<1/\sqrt{3\,}$, meaning that after the bounce, when the universe starts expanding again and the wall moves towards the right, there will be a second Kasner reflection bringing $r$ to $r^{(2)}>1/\sqrt{3}$ or $r^{(2)}<-1/\sqrt{3}$ (for positive or negative $\overline{\,\pi_-}$ respectively).
\end{itemize}

\begin{figure}
    \centering
    \includegraphics[width=0.9\linewidth]{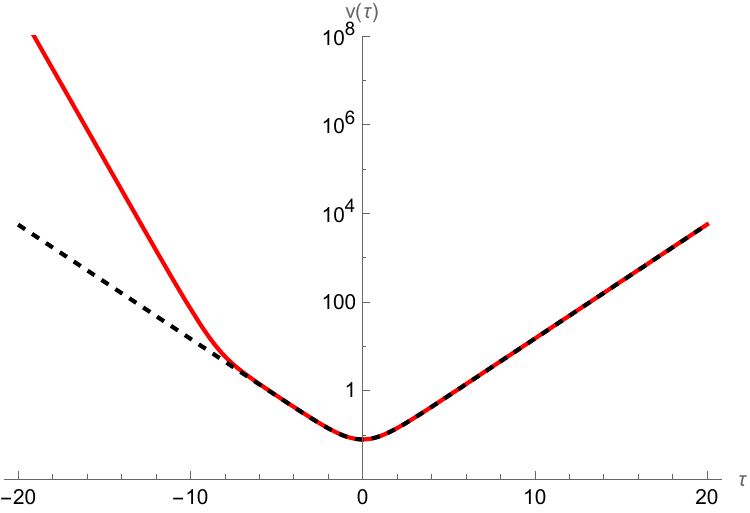}
    \includegraphics[width=0.9\linewidth]{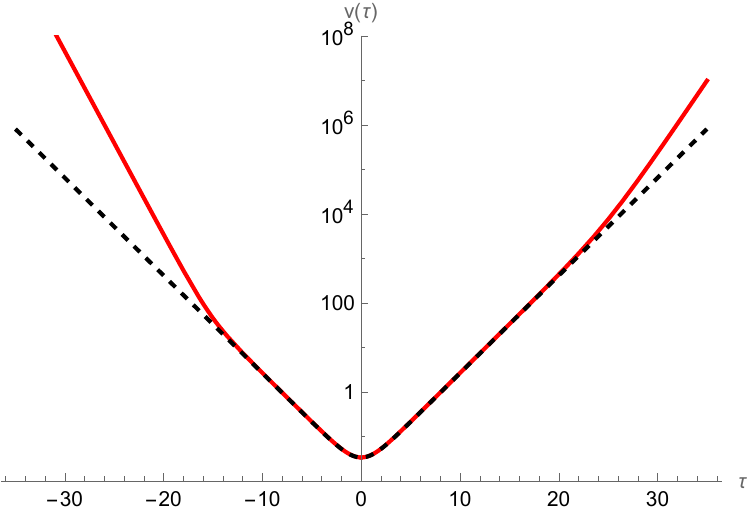}
    \caption{Evolution of the volume $v$ as function of harmonic time $\tau$ in the deformed Bianchi II model (red lines) for the case of one reflection (top) or two (bottom), compared with the simple bouncing Bianchi I (black dashed lines).}
    \label{fig:modBianchiII}
\end{figure}

Fig.~\ref{fig:modBianchiII} shows  two of the three scenarios (the first and second ones are essentially the same but time-reversed), obtained through numerical integration, compared with the bouncing Bianchi I in order to better visualise the Kasner reflections.

Let us now consider deformed anisotropies with ${f\neq1}$. The limiting values for $k_3$ that decide which scenario takes place are not changed, because the Kasner map \eqref{classKasnermap} is unaffected; however, for $f>1$ the allowed ranges for the indices are much wider, according to the deformed second Kasner relation \eqref{deformedKasner2}. In particular, the indices can now be large in absolute value, and the ranges for the three scenarios become
\begin{itemize}
    \item $k_3^{(0)}>0$: only one Kasner reflection after the bounce;
    \item $-2/7<k_3^{(0)}<0$: two Kasner reflections, one before and one after the bounce;
    \item $k_3^{(0)}<-2/7$: only one Kasner reflection before the bounce.
\end{itemize}
The only difference is that the overall upper and lower limits for Kasner indices are changed compared to the usual values of $1$ and $-1/3$, depending on what values $f$ can take.

As mentioned above, in the numerical calculations we used for the anisotropies the same cut-off algebra used for the volume, which here takes the form
\begin{equation}
    f(\pi_\text{tot})=\sqrt{1-\mu_\beta^2(\pi_+^2+\pi_-^2)\,}\,.
    \label{fchoice}
\end{equation}
First of all, since we always have $f\le 1$, the allowed upper and lower limits for Kasner indices are closer to zero than in the standard case. Furthermore, as we already know from deforming the volume, this algebra implements a cut-off for the total momentum, $\pi_\text{tot}^2<1/\mu_\beta^2$. Therefore the two momenta $\pi_\pm$ are bound to the circle of radius $1/\mu_\beta$, meaning that for any given value of the constant $\overline{\,\pi_-}$ (whose absolute value is also constrained to be smaller than that radius) there will be a bound on the allowed values of the ratio $r$ given by
\begin{equation}
    \abs{r}<r_\text{max}=\sqrt{\frac{1}{\mu_\beta^2\overline{\,\pi_-}^2\,}-1\,}\,.
    \label{rupperlimit}
\end{equation}
Finally, the function $f$ also changes the relation between the indices and the anisotropic momenta, as shown in Eq.~\eqref{pitoKasnerindicesdeformed}. All considered, to see how $f$ affects the ranges of the initial conditions on the ratio $r$, we have to rely on numerical calculations. Using a specific choice for $f$ such as (\ref{fchoice}), the limiting values for $k_3$ that decide how many reflections take place can be translated into their corresponding values in $r$. In any case, the saturation of condition \eqref{rupperlimit} at $r=\pm r_\text{max}$ implies $f=0$, and therefore ${k_3(\pm r_\text{max})=1/3}$.

\begin{figure}
    \centering
    \includegraphics[width=0.9\linewidth]{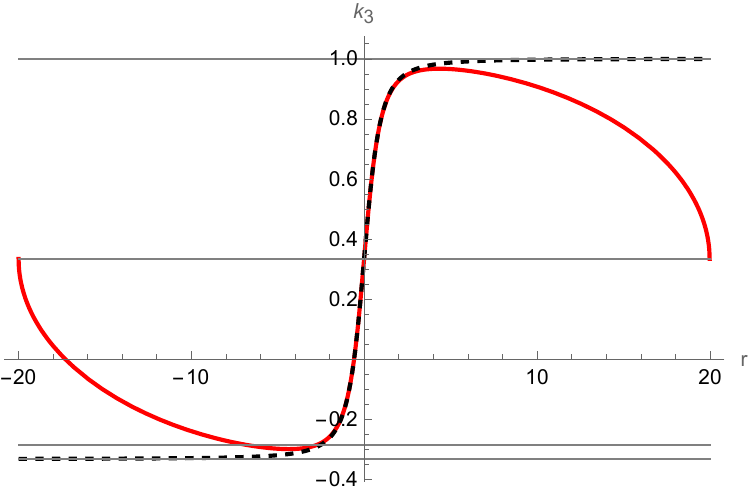}
    \includegraphics[width=0.9\linewidth]{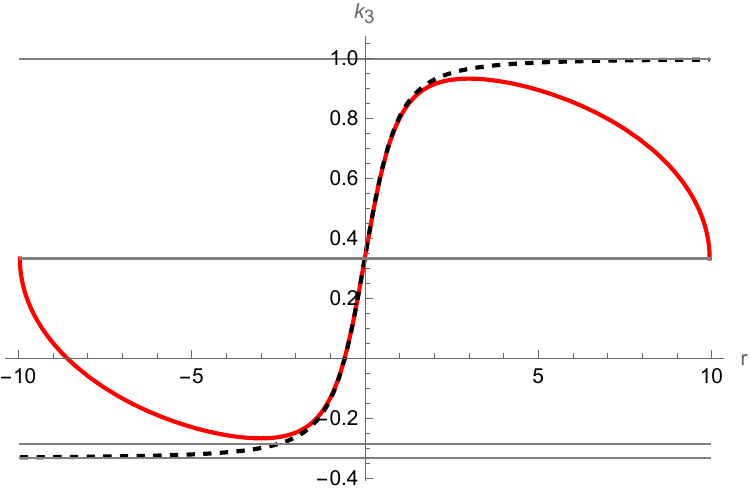}
    \caption{The deformed Kasner index $k_3$ (red continuous line) as function of the ratio $r$ in the two cases of $\mu_\beta=1/20$ (top) and $\mu_\beta=1/10$ (bottom), compared with the corresponding classical function (black dashed lines). The thin grey horizontal lines correspond to the values $1$, $1/3$, $-2/7$, and $-1/3$. Notice in particular that the values $k_3<-2/7$ needed to only have one reflection before the bounce are only obtained for a small range of $r$ values (top) or not reached at all (bottom).}
    \label{fig:modk3}
\end{figure}

Here we focus on two specific numerical examples. First we consider $\mu_\beta=1/20$ and ${\overline{\,\pi_-}=1>0}$; then we have $r_\text{max}=19.975$ and, recalling that $1/\sqrt{3\,}\approx0.577$ and $13/(3\sqrt{3})\approx2.502$, the three scenarios are
\begin{itemize}
    \item ${-0.579<r^{(0)}<r_\text{max}}$ or ${-r_\text{max}<r^{(0)}<-17.28}$: only one Kasner reflection after the bounce;
    \item For either $-17.28<r^{(0)}<-6.835$ or ${-2.717<r^{(0)}<-0.579}$, we have two Kasner reflections, one before and one after the bounce;
    \item ${-6.835<r^{(0)}<-2.717}$: only one Kasner reflection before the bounce.
\end{itemize}
It is evident how the values $0.579$ and $2.717$ are deformations of $1/\sqrt{3\,}$ and $13/(3\sqrt{3})$ respectively. However, we see that new intervals appear: the condition $k_3^{(0)}>0$ that results in only one reflection after the bounce is satisfied not just for $r^{(0)}>-1/\sqrt{3\,}$, but also for a small interval near the negative lower limit for $r$. The new limiting value $-6.835$ appears because the smaller the function $f$ is, i.e., the bigger $\mu_\beta$ is, the closer the function $k_3(r)$ will flatten towards the value of $1/3$ (as mentioned above, when $f$ is exactly zero, all indices will flatten to this value) and therefore it will be harder and harder to achieve the negative values necessary for that scenario. Indeed, let us show a second example with a bigger value for the deformation parameter. For $\mu_\beta=1/10$ we have ${r_\text{max}=9.950}$ and
\begin{itemize}
    \item ${-0.583<r^{(0)}<r_\text{max}}$ or ${-r_\text{max}<r^{(0)}<-8.583}$: only one Kasner reflection after the bounce;
    \item ${-8.583<r^{(0)}<-0.583}$: two Kasner reflections, one before and one after the bounce;
    \item there are no values of $r^{(0)}$ such that the third scenario with only one Kasner reflection before the bounce happens.
\end{itemize}
Again we have a new small interval near the lower bound which has the same properties as the bigger interval that is the deformation of $r>-1/\sqrt{3\,}$; however now the value of $f$ is so small that it is not possible to obtain ${k_3^{(0)}<-2/7}$. Fig.~\ref{fig:modk3} shows the relevant Kasner index as a function of the ratio between momenta in the case of these two numerical examples, to better visualise the various ranges for $r$ and show how to obtain them from the limits on $k_3$. Besides showing the different values of $r_\text{max}$, it is clear how in the case $\mu_\beta=1/20$ the function does intersect (and go below) the value $-2/7$, while in the case $\mu_\beta=1/10$ it does not, being more flattened towards $1/3$.

The appearance of smaller intervals close to the opposite lower or upper limit is indeed due to the fact that, when $r$ is near $\pm r_\text{max}$, the function $f$ gets close to zero. This is similar to what happens with lattices where quantities acquire some periodic properties and there are additional low-energy excitations at large momentum. This again suggests that the cut-off algebra we have implemented for the volume and anisotropies might be related to some type of lattice quantisation of gravity.

\subsection{Anisotropy time}

As in the case of Bianchi I, it is insightful to convert the Friedmann equation from synchronous time (\ref{DefBianchiIIFried}) to a relational notion of time, in order to characterise observables of the system. Again, the obvious candidate for such a clock is given by anisotropy variables $\beta_\pm$. However, differently from  the Bianchi I case, in a Bianchi II model the two anisotropies are not on the same footing any more: given the Kasner reflection off of the potential wall, it is clear that the variable $\beta_+$ is not monotonic  with respect to either $t$ or $\tau$. Therefore in this case it is better to use $\beta_-$, which is still monotonic, although slightly modified by the presence of $f$ (which depends on time through $\pi_+$) and by noncommutative effects.

In a classical Bianchi II model, the corresponding Friedmann equation is
\begin{equation}
    \left(\frac{1}{3v}\,\dv{v}{\beta_-}\right)^2=\frac{\pi_\text{tot}^2+12v^2U}{\pi_-^2}\,.
\end{equation}
In the deformed case instead we have
\begin{equation}
    \left(\frac{1}{3v}\,\dv{v}{\beta_-}\right)^2=\frac{\left(\pi_\text{tot}^2+12v^2U\right)\left(1-\frac{\rho_a+U}{\rho_\mu}\right)}{\Bigl(\pi_-f(\pi_\text{tot})+\pi_\text{tot}f'(\pi_\text{tot})J\frac{U}{\rho_a}\Bigr)^2}\,,
\end{equation}
where again we see both the LQC-like correction coming from the volume deformation and the correction at the denominator coming from $f(\pi_\text{tot})\neq1$. This is in contrast to the Friedmann equation in synchronous time where $f$ does not appear, again in analogy with what we had found for Bianchi I (cf.~(\ref{relationalFriedmann})).

\section{Conclusions}
\label{Sec:concl}

We have studied a modification of the classical Bianchi I and Bianchi II models in vacuum, obtained from deforming the Poisson brackets between Misner-like metric components and their respective momenta. For the volume, our deformation was fixed by requiring it to reproduce a bounce similar to the one seen in isotropic LQC, and we did indeed find such a bounce also in the anisotropic case. It seems to be a natural minimal requirement to ask that isotropic cosmology arises as a special case of anisotropic cosmology, and one would expect this to be true in any putative theory of quantum gravity. Conversely, this also means that the deformation introduced for anisotropic degrees of freedom, here represented by the Misner variables $\beta_\pm$, is not fixed by any fundamental principle in our approach. In much of the analysis, this deformation was left as a free function $f$ which enters directly in the effective Friedmann equation. We saw that this deformation, which acts as a simple constant or (smoothly) piecewise constant rescaling, remains relevant at late times and low energies in the post-bounce phase even though its origin is assumed in the quantum realm. This might help in constraining some forms through observations.

Our approach is a general framework in which different types of potential quantum effects could be included. In the more specific case where the anisotropy variables are deformed by using the same cut-off algebra used for the volume, we found indications that this algebra mimics the effects of a discretisation of these variables, similar to the discretisation of the volume used in LQC. Noting that this specific feature is different from a minimal length or volume approach, it would be very interesting to see whether such a ``discrete anisotropy'' can be related to a quantum gravity model. One advantage in our approach is that effective dynamical equations and Friedmann equations are obtained straightforwardly, which can be contrasted, e.g., with the rather complicated form of Bianchi models in LQC. Some features of the deformed Bianchi II model, such as reflection laws for the Kasner coefficients and the existence of different solutions in which either one or two Kasner reflections occur, were found to be in agreement with LQC, again pointing towards a general applicability of the approach used here.

An interesting extension of our work would be to the Bianchi IX model, which plays the particularly important role of representing the generic approach to spacelike singularities in general relativity. Some preliminary work on DCRs in the Bianchi IX model has been done based on deforming the anisotropy variables alone \cite{Segreto:2024gcg,Barca:2024goe}, i.e., without including a deformation that would cause a bounce. Including also the volume deformation would make this another test case for the study of more complicated or realistic bouncing cosmologies.

It would also be interesting to investigate a possible connection of the DCR approach to settings in which quantum effects arise from particular features of the quantisation. For instance, resolution of singularities in a Bianchi I model can also be achieved in affine quantisation \cite{Bergeron:2015jpa,Malkiewicz:2019azw}, which effectively leads to additional terms in the Hamiltonian. It seems plausible that such effects could also be mimicked via DCRs, further extending their applicability.

Regarding possible connection with observations, it might also be interesting to compute how a primordial power spectrum of perturbations (e.g., within inflation) might be modified by a deformed background containing not just a LQC-like bounce, but also deformed anisotropies. Such calculations could be compared with power spectra derived from both standard Bianchi models \cite{MalkiewiczBianchiCMB} and bouncing isotropic models \cite{Agullo:2013ai}.

\acknowledgments
The authors thank S. Segreto and S. F. Uria for useful insights and discussions. The work of SG is funded by the Royal Society through the University Research Fellowship Renewal URF$\backslash$R$\backslash$221005.

\section*{Data availability}
The data that support the findings of this article are openly available \cite{arXivversion}. 

\bibliographystyle{apsrev4-2}
\bibliography{DeformedBianchi.bib}

\begin{thebibliography}{43}%
\makeatletter
\providecommand \@ifxundefined [1]{%
 \@ifx{#1\undefined}
}%
\providecommand \@ifnum [1]{%
 \ifnum #1\expandafter \@firstoftwo
 \else \expandafter \@secondoftwo
 \fi
}%
\providecommand \@ifx [1]{%
 \ifx #1\expandafter \@firstoftwo
 \else \expandafter \@secondoftwo
 \fi
}%
\providecommand \natexlab [1]{#1}%
\providecommand \enquote  [1]{``#1''}%
\providecommand \bibnamefont  [1]{#1}%
\providecommand \bibfnamefont [1]{#1}%
\providecommand \citenamefont [1]{#1}%
\providecommand \href@noop [0]{\@secondoftwo}%
\providecommand \href [0]{\begingroup \@sanitize@url \@href}%
\providecommand \@href[1]{\@@startlink{#1}\@@href}%
\providecommand \@@href[1]{\endgroup#1\@@endlink}%
\providecommand \@sanitize@url [0]{\catcode `\\12\catcode `\$12\catcode
  `\&12\catcode `\#12\catcode `\^12\catcode `\_12\catcode `\%12\relax}%
\providecommand \@@startlink[1]{}%
\providecommand \@@endlink[0]{}%
\providecommand \url  [0]{\begingroup\@sanitize@url \@url }%
\providecommand \@url [1]{\endgroup\@href {#1}{\urlprefix }}%
\providecommand \urlprefix  [0]{URL }%
\providecommand \Eprint [0]{\href }%
\providecommand \doibase [0]{https://doi.org/}%
\providecommand \selectlanguage [0]{\@gobble}%
\providecommand \bibinfo  [0]{\@secondoftwo}%
\providecommand \bibfield  [0]{\@secondoftwo}%
\providecommand \translation [1]{[#1]}%
\providecommand \BibitemOpen [0]{}%
\providecommand \bibitemStop [0]{}%
\providecommand \bibitemNoStop [0]{.\EOS\space}%
\providecommand \EOS [0]{\spacefactor3000\relax}%
\providecommand \BibitemShut  [1]{\csname bibitem#1\endcsname}%
\let\auto@bib@innerbib\@empty
\bibitem [{\citenamefont {Hawking}\ and\ \citenamefont
  {Penrose}(1970)}]{SingularityTheorems1}%
  \BibitemOpen
  \bibfield  {author} {\bibinfo {author} {\bibfnamefont {S.~W.}\ \bibnamefont
  {Hawking}}\ and\ \bibinfo {author} {\bibfnamefont {R.}~\bibnamefont
  {Penrose}},\ }\href {https://doi.org/10.1098/rspa.1970.0021} {\bibfield
  {journal} {\bibinfo  {journal} {Proceedings of the Royal Society of London
  A}\ }\textbf {\bibinfo {volume} {314}},\ \bibinfo {pages} {529} (\bibinfo
  {year} {1970})}\BibitemShut {NoStop}%
\bibitem [{\citenamefont {Hawking}\ and\ \citenamefont
  {Penrose}(1996)}]{SingularityTheorems2}%
  \BibitemOpen
  \bibfield  {author} {\bibinfo {author} {\bibfnamefont {S.~W.}\ \bibnamefont
  {Hawking}}\ and\ \bibinfo {author} {\bibfnamefont {R.}~\bibnamefont
  {Penrose}},\ }\href
  {https://press.princeton.edu/books/paperback/9780691168449/the-nature-of-space-and-time}
  {\emph {\bibinfo {title} {The {N}ature of {S}pace and {T}ime}}}\ (\bibinfo
  {publisher} {Princeton University Press},\ \bibinfo {address} {Princeton, New
  Jersey},\ \bibinfo {year} {1996})\BibitemShut {NoStop}%
\bibitem [{\citenamefont {Bojowald}(2005)}]{BojowaldOriginalLQC}%
  \BibitemOpen
  \bibfield  {author} {\bibinfo {author} {\bibfnamefont {M.}~\bibnamefont
  {Bojowald}},\ }\href {https://doi.org/10.12942/lrr-2005-11} {\bibfield
  {journal} {\bibinfo  {journal} {Liv. Rev. Rel.}\ }\textbf {\bibinfo {volume}
  {8}},\ \bibinfo {pages} {11} (\bibinfo {year} {2005})},\ \Eprint
  {https://arxiv.org/abs/gr-qc/0601085} {arXiv:gr-qc/0601085} \BibitemShut
  {NoStop}%
\bibitem [{\citenamefont {Ashtekar}\ and\ \citenamefont
  {Singh}(2011)}]{LQC2011Review}%
  \BibitemOpen
  \bibfield  {author} {\bibinfo {author} {\bibfnamefont {A.}~\bibnamefont
  {Ashtekar}}\ and\ \bibinfo {author} {\bibfnamefont {P.}~\bibnamefont
  {Singh}},\ }\href {https://doi.org/10.1088/0264-9381/28/21/213001} {\bibfield
   {journal} {\bibinfo  {journal} {Class. Quant. Grav.}\ }\textbf {\bibinfo
  {volume} {28}},\ \bibinfo {pages} {213001} (\bibinfo {year} {2011})},\
  \Eprint {https://arxiv.org/abs/1108.0893} {arXiv:1108.0893 [gr-qc]}
  \BibitemShut {NoStop}%
\bibitem [{\citenamefont {Thiemann}(2007)}]{Thiemann_2007}%
  \BibitemOpen
  \bibfield  {author} {\bibinfo {author} {\bibfnamefont {T.}~\bibnamefont
  {Thiemann}},\ }\href {https://doi.org/10.1017/CBO9780511755682} {\emph
  {\bibinfo {title} {Modern Canonical Quantum General Relativity}}}\ (\bibinfo
  {publisher} {Cambridge University Press},\ \bibinfo {address} {Cambridge},\
  \bibinfo {year} {2007})\BibitemShut {NoStop}%
\bibitem [{\citenamefont {Ashtekar}\ and\ \citenamefont
  {Pullin}(2017)}]{LQGreview}%
  \BibitemOpen
  \bibfield  {author} {\bibinfo {author} {\bibfnamefont {A.}~\bibnamefont
  {Ashtekar}}\ and\ \bibinfo {author} {\bibfnamefont {J.}~\bibnamefont
  {Pullin}},\ }\href {https://doi.org/10.1142/10445} {\emph {\bibinfo {title}
  {Loop {Q}uantum {G}ravity: {T}he {F}irst 30 {Y}ears}}}\ (\bibinfo
  {publisher} {World Scientific},\ \bibinfo {address} {Singapore},\ \bibinfo
  {year} {2017})\BibitemShut {NoStop}%
\bibitem [{\citenamefont {Assanioussi}\ \emph {et~al.}(2018)\citenamefont
  {Assanioussi}, \citenamefont {Dapor}, \citenamefont {Liegener},\ and\
  \citenamefont {Paw{\l}owski}}]{Assanioussi:2018hee}%
  \BibitemOpen
  \bibfield  {author} {\bibinfo {author} {\bibfnamefont {M.}~\bibnamefont
  {Assanioussi}}, \bibinfo {author} {\bibfnamefont {A.}~\bibnamefont {Dapor}},
  \bibinfo {author} {\bibfnamefont {K.}~\bibnamefont {Liegener}},\ and\
  \bibinfo {author} {\bibfnamefont {T.}~\bibnamefont {Paw{\l}owski}},\ }\href
  {https://doi.org/10.1103/PhysRevLett.121.081303} {\bibfield  {journal}
  {\bibinfo  {journal} {Phys. Rev. Lett.}\ }\textbf {\bibinfo {volume} {121}},\
  \bibinfo {pages} {081303} (\bibinfo {year} {2018})},\ \Eprint
  {https://arxiv.org/abs/1801.00768} {arXiv:1801.00768 [gr-qc]} \BibitemShut
  {NoStop}%
\bibitem [{\citenamefont {Liegener}\ and\ \citenamefont
  {Singh}(2020)}]{Liegener:2019zgw}%
  \BibitemOpen
  \bibfield  {author} {\bibinfo {author} {\bibfnamefont {K.}~\bibnamefont
  {Liegener}}\ and\ \bibinfo {author} {\bibfnamefont {P.}~\bibnamefont
  {Singh}},\ }\href {https://doi.org/10.1088/1361-6382/ab7962} {\bibfield
  {journal} {\bibinfo  {journal} {Class. Quant. Grav.}\ }\textbf {\bibinfo
  {volume} {37}},\ \bibinfo {pages} {085015} (\bibinfo {year} {2020})},\
  \Eprint {https://arxiv.org/abs/1906.02759} {arXiv:1906.02759 [gr-qc]}
  \BibitemShut {NoStop}%
\bibitem [{\citenamefont {Dapor}\ \emph {et~al.}(2019)\citenamefont {Dapor},
  \citenamefont {Liegener},\ and\ \citenamefont
  {Paw{\l}owski}}]{Dapor:2019mil}%
  \BibitemOpen
  \bibfield  {author} {\bibinfo {author} {\bibfnamefont {A.}~\bibnamefont
  {Dapor}}, \bibinfo {author} {\bibfnamefont {K.}~\bibnamefont {Liegener}},\
  and\ \bibinfo {author} {\bibfnamefont {T.}~\bibnamefont {Paw{\l}owski}},\
  }\href {https://doi.org/10.1103/PhysRevD.100.106016} {\bibfield  {journal}
  {\bibinfo  {journal} {Phys. Rev. D}\ }\textbf {\bibinfo {volume} {100}},\
  \bibinfo {pages} {106016} (\bibinfo {year} {2019})},\ \Eprint
  {https://arxiv.org/abs/1910.04710} {arXiv:1910.04710 [gr-qc]} \BibitemShut
  {NoStop}%
\bibitem [{\citenamefont {Oriti}\ \emph {et~al.}(2016)\citenamefont {Oriti},
  \citenamefont {Sindoni},\ and\ \citenamefont {Wilson-Ewing}}]{Oriti:2016qtz}%
  \BibitemOpen
  \bibfield  {author} {\bibinfo {author} {\bibfnamefont {D.}~\bibnamefont
  {Oriti}}, \bibinfo {author} {\bibfnamefont {L.}~\bibnamefont {Sindoni}},\
  and\ \bibinfo {author} {\bibfnamefont {E.}~\bibnamefont {Wilson-Ewing}},\
  }\href {https://doi.org/10.1088/0264-9381/33/22/224001} {\bibfield  {journal}
  {\bibinfo  {journal} {Class. Quant. Grav.}\ }\textbf {\bibinfo {volume}
  {33}},\ \bibinfo {pages} {224001} (\bibinfo {year} {2016})},\ \Eprint
  {https://arxiv.org/abs/1602.05881} {arXiv:1602.05881 [gr-qc]} \BibitemShut
  {NoStop}%
\bibitem [{\citenamefont {Gielen}\ and\ \citenamefont
  {Polaczek}(2020)}]{Gielen:2019kae}%
  \BibitemOpen
  \bibfield  {author} {\bibinfo {author} {\bibfnamefont {S.}~\bibnamefont
  {Gielen}}\ and\ \bibinfo {author} {\bibfnamefont {A.}~\bibnamefont
  {Polaczek}},\ }\href {https://doi.org/10.1088/1361-6382/ab8f67} {\bibfield
  {journal} {\bibinfo  {journal} {Class. Quant. Grav.}\ }\textbf {\bibinfo
  {volume} {37}},\ \bibinfo {pages} {165004} (\bibinfo {year} {2020})},\
  \Eprint {https://arxiv.org/abs/1912.06143} {arXiv:1912.06143 [gr-qc]}
  \BibitemShut {NoStop}%
\bibitem [{\citenamefont {Bojowald}(2020)}]{LQCproblems1}%
  \BibitemOpen
  \bibfield  {author} {\bibinfo {author} {\bibfnamefont {M.}~\bibnamefont
  {Bojowald}},\ }\href {https://doi.org/10.3390/universe6030036} {\bibfield
  {journal} {\bibinfo  {journal} {Universe}\ }\textbf {\bibinfo {volume} {6}},\
  \bibinfo {pages} {36} (\bibinfo {year} {2020})},\ \Eprint
  {https://arxiv.org/abs/2002.05703} {arXiv:2002.05703 [gr-qc]} \BibitemShut
  {NoStop}%
\bibitem [{\citenamefont {Hossenfelder}(2013)}]{Hossenfelder}%
  \BibitemOpen
  \bibfield  {author} {\bibinfo {author} {\bibfnamefont {S.}~\bibnamefont
  {Hossenfelder}},\ }\href {https://doi.org/10.12942/lrr-2013-2} {\bibfield
  {journal} {\bibinfo  {journal} {Liv. Rev. Rel.}\ }\textbf {\bibinfo {volume}
  {16}},\ \bibinfo {pages} {2} (\bibinfo {year} {2013})},\ \Eprint
  {https://arxiv.org/abs/1203.6191} {arXiv:1203.6191} \BibitemShut {NoStop}%
\bibitem [{\citenamefont {Kempf}\ \emph {et~al.}(1995)\citenamefont {Kempf},
  \citenamefont {Mangano},\ and\ \citenamefont {Mann}}]{KMM}%
  \BibitemOpen
  \bibfield  {author} {\bibinfo {author} {\bibfnamefont {A.}~\bibnamefont
  {Kempf}}, \bibinfo {author} {\bibfnamefont {G.}~\bibnamefont {Mangano}},\
  and\ \bibinfo {author} {\bibfnamefont {R.~B.}\ \bibnamefont {Mann}},\ }\href
  {https://doi.org/10.1103/physrevd.52.1108} {\bibfield  {journal} {\bibinfo
  {journal} {Phys. Rev. D}\ }\textbf {\bibinfo {volume} {52}},\ \bibinfo
  {pages} {1108} (\bibinfo {year} {1995})},\ \Eprint
  {https://arxiv.org/abs/hep-th/9412167} {arXiv:hep-th/9412167} \BibitemShut
  {NoStop}%
\bibitem [{\citenamefont {Maggiore}(1993)}]{Maggiore93}%
  \BibitemOpen
  \bibfield  {author} {\bibinfo {author} {\bibfnamefont {M.}~\bibnamefont
  {Maggiore}},\ }\href {https://doi.org/10.1016/0370-2693(93)91401-8}
  {\bibfield  {journal} {\bibinfo  {journal} {Phys. Lett. B}\ }\textbf
  {\bibinfo {volume} {304}},\ \bibinfo {pages} {65} (\bibinfo {year} {1993})},\
  \Eprint {https://arxiv.org/abs/hep-th/9301067} {arXiv:hep-th/9301067}
  \BibitemShut {NoStop}%
\bibitem [{\citenamefont {Scardigli}(1999)}]{ScardigliGUP}%
  \BibitemOpen
  \bibfield  {author} {\bibinfo {author} {\bibfnamefont {F.}~\bibnamefont
  {Scardigli}},\ }\href {https://doi.org/10.1016/S0370-2693(99)00167-7}
  {\bibfield  {journal} {\bibinfo  {journal} {Phys. Lett. B}\ }\textbf
  {\bibinfo {volume} {452}},\ \bibinfo {pages} {39} (\bibinfo {year} {1999})},\
  \Eprint {https://arxiv.org/abs/hep-th/9904025} {arXiv:hep-th/9904025}
  \BibitemShut {NoStop}%
\bibitem [{\citenamefont {Amati}\ \emph {et~al.}(1989)\citenamefont {Amati},
  \citenamefont {Ciafaloni},\ and\ \citenamefont {Veneziano}}]{ST1}%
  \BibitemOpen
  \bibfield  {author} {\bibinfo {author} {\bibfnamefont {D.}~\bibnamefont
  {Amati}}, \bibinfo {author} {\bibfnamefont {M.}~\bibnamefont {Ciafaloni}},\
  and\ \bibinfo {author} {\bibfnamefont {G.}~\bibnamefont {Veneziano}},\ }\href
  {https://doi.org/10.1016/0370-2693(89)91366-X} {\bibfield  {journal}
  {\bibinfo  {journal} {Phys. Lett. B}\ }\textbf {\bibinfo {volume} {216}},\
  \bibinfo {pages} {41} (\bibinfo {year} {1989})}\BibitemShut {NoStop}%
\bibitem [{\citenamefont {Amelino-Camelia}\ \emph {et~al.}(1997)\citenamefont
  {Amelino-Camelia}, \citenamefont {Mavromatos}, \citenamefont {Ellis},\ and\
  \citenamefont {Nanopoulos}}]{ST2}%
  \BibitemOpen
  \bibfield  {author} {\bibinfo {author} {\bibfnamefont {G.}~\bibnamefont
  {Amelino-Camelia}}, \bibinfo {author} {\bibfnamefont {N.~E.}\ \bibnamefont
  {Mavromatos}}, \bibinfo {author} {\bibfnamefont {J.}~\bibnamefont {Ellis}},\
  and\ \bibinfo {author} {\bibfnamefont {D.~V.}\ \bibnamefont {Nanopoulos}},\
  }\href {https://doi.org/10.1142/s0217732397002077} {\bibfield  {journal}
  {\bibinfo  {journal} {Mod. Phys. Lett. A}\ }\textbf {\bibinfo {volume}
  {12}},\ \bibinfo {pages} {2029} (\bibinfo {year} {1997})},\ \Eprint
  {https://arxiv.org/abs/hep-th/9701144} {arXiv:hep-th/9701144} \BibitemShut
  {NoStop}%
\bibitem [{\citenamefont {Battisti}(2009)}]{Battisti}%
  \BibitemOpen
  \bibfield  {author} {\bibinfo {author} {\bibfnamefont {M.~V.}\ \bibnamefont
  {Battisti}},\ }\href {https://doi.org/10.1103/PhysRevD.79.083506} {\bibfield
  {journal} {\bibinfo  {journal} {Phys. Rev. D}\ }\textbf {\bibinfo {volume}
  {79}},\ \bibinfo {pages} {083506} (\bibinfo {year} {2009})}\BibitemShut
  {NoStop}%
\bibitem [{\citenamefont {Barca}\ \emph {et~al.}(2022)\citenamefont {Barca},
  \citenamefont {Giovannetti},\ and\ \citenamefont {Montani}}]{IJGMMP}%
  \BibitemOpen
  \bibfield  {author} {\bibinfo {author} {\bibfnamefont {G.}~\bibnamefont
  {Barca}}, \bibinfo {author} {\bibfnamefont {E.}~\bibnamefont {Giovannetti}},\
  and\ \bibinfo {author} {\bibfnamefont {G.}~\bibnamefont {Montani}},\ }\href
  {https://doi.org/10.1142/s0219887822500979} {\bibfield  {journal} {\bibinfo
  {journal} {Int. J. Geom. Meth. Mod. Phys.}\ }\textbf {\bibinfo {volume}
  {19}},\ \bibinfo {pages} {2250097} (\bibinfo {year} {2022})},\ \Eprint
  {https://arxiv.org/abs/2112.08905} {arXiv:2112.08905} \BibitemShut {NoStop}%
\bibitem [{\citenamefont {Barca}\ \emph {et~al.}(2023)\citenamefont {Barca},
  \citenamefont {Montani},\ and\ \citenamefont {Melchiorri}}]{BarcaEU}%
  \BibitemOpen
  \bibfield  {author} {\bibinfo {author} {\bibfnamefont {G.}~\bibnamefont
  {Barca}}, \bibinfo {author} {\bibfnamefont {G.}~\bibnamefont {Montani}},\
  and\ \bibinfo {author} {\bibfnamefont {A.}~\bibnamefont {Melchiorri}},\
  }\href {https://doi.org/10.1103/PhysRevD.108.063505} {\bibfield  {journal}
  {\bibinfo  {journal} {Phys. Rev. D}\ }\textbf {\bibinfo {volume} {108}},\
  \bibinfo {pages} {063505} (\bibinfo {year} {2023})},\ \Eprint
  {https://arxiv.org/abs/2302.01173} {arXiv:2302.01173 [gr-qc]} \BibitemShut
  {NoStop}%
\bibitem [{\citenamefont {Fadel}\ and\ \citenamefont
  {Maggiore}(2022)}]{Maggiore2021}%
  \BibitemOpen
  \bibfield  {author} {\bibinfo {author} {\bibfnamefont {M.}~\bibnamefont
  {Fadel}}\ and\ \bibinfo {author} {\bibfnamefont {M.}~\bibnamefont
  {Maggiore}},\ }\href {https://doi.org/10.1103/physrevd.105.106017} {\bibfield
   {journal} {\bibinfo  {journal} {Phys. Rev. D}\ }\textbf {\bibinfo {volume}
  {105}},\ \bibinfo {pages} {106017} (\bibinfo {year} {2022})},\ \Eprint
  {https://arxiv.org/abs/2112.09034} {arXiv:2112.09034} \BibitemShut {NoStop}%
\bibitem [{\citenamefont {Bruno}\ \emph {et~al.}(2024)\citenamefont {Bruno},
  \citenamefont {Segreto},\ and\ \citenamefont
  {Montani}}]{SebyMatteoSemiclassicalGUP}%
  \BibitemOpen
  \bibfield  {author} {\bibinfo {author} {\bibfnamefont {M.}~\bibnamefont
  {Bruno}}, \bibinfo {author} {\bibfnamefont {S.}~\bibnamefont {Segreto}},\
  and\ \bibinfo {author} {\bibfnamefont {G.}~\bibnamefont {Montani}},\ }\href
  {https://doi.org/10.1016/j.nuclphysb.2024.116739} {\bibfield  {journal}
  {\bibinfo  {journal} {Nucl. Phys. B}\ }\textbf {\bibinfo {volume} {1009}},\
  \bibinfo {pages} {116739} (\bibinfo {year} {2024})},\ \Eprint
  {https://arxiv.org/abs/2407.17408} {arXiv:2407.17408 [math-ph]} \BibitemShut
  {NoStop}%
\bibitem [{\citenamefont {Taveras}(2008)}]{Taveras:2008ke}%
  \BibitemOpen
  \bibfield  {author} {\bibinfo {author} {\bibfnamefont {V.}~\bibnamefont
  {Taveras}},\ }\href {https://doi.org/10.1103/PhysRevD.78.064072} {\bibfield
  {journal} {\bibinfo  {journal} {Phys. Rev. D}\ }\textbf {\bibinfo {volume}
  {78}},\ \bibinfo {pages} {064072} (\bibinfo {year} {2008})},\ \Eprint
  {https://arxiv.org/abs/0807.3325} {arXiv:0807.3325 [gr-qc]} \BibitemShut
  {NoStop}%
\bibitem [{\citenamefont {Wilson-Ewing}(2018)}]{WilsonEwingBianchiLQC}%
  \BibitemOpen
  \bibfield  {author} {\bibinfo {author} {\bibfnamefont {E.}~\bibnamefont
  {Wilson-Ewing}},\ }\href {https://doi.org/10.1088/1361-6382/aaab8b}
  {\bibfield  {journal} {\bibinfo  {journal} {Class. Quant. Grav.}\ }\textbf
  {\bibinfo {volume} {35}},\ \bibinfo {pages} {065005} (\bibinfo {year}
  {2018})},\ \Eprint {https://arxiv.org/abs/1711.10943} {arXiv:1711.10943}
  \BibitemShut {NoStop}%
\bibitem [{\citenamefont {Misner}\ \emph {et~al.}(1973)\citenamefont {Misner},
  \citenamefont {Thorne},\ and\ \citenamefont {Wheeler}}]{MisnerGravitation}%
  \BibitemOpen
  \bibfield  {author} {\bibinfo {author} {\bibfnamefont {C.~W.}\ \bibnamefont
  {Misner}}, \bibinfo {author} {\bibfnamefont {K.~S.}\ \bibnamefont {Thorne}},\
  and\ \bibinfo {author} {\bibfnamefont {J.~A.}\ \bibnamefont {Wheeler}},\
  }\href@noop {} {\emph {\bibinfo {title} {Gravitation}}}\ (\bibinfo
  {publisher} {W. H. Freeman},\ \bibinfo {address} {San Francisco,
  California},\ \bibinfo {year} {1973})\BibitemShut {NoStop}%
\bibitem [{\citenamefont {Montani}\ \emph {et~al.}(2011)\citenamefont
  {Montani}, \citenamefont {Battisti}, \citenamefont {Benini},\ and\
  \citenamefont {Imponente}}]{PrimordialCosmology}%
  \BibitemOpen
  \bibfield  {author} {\bibinfo {author} {\bibfnamefont {G.}~\bibnamefont
  {Montani}}, \bibinfo {author} {\bibfnamefont {M.~V.}\ \bibnamefont
  {Battisti}}, \bibinfo {author} {\bibfnamefont {R.}~\bibnamefont {Benini}},\
  and\ \bibinfo {author} {\bibfnamefont {G.}~\bibnamefont {Imponente}},\
  }\href@noop {} {\emph {\bibinfo {title} {Primordial Cosmology}}}\ (\bibinfo
  {publisher} {World Scientific},\ \bibinfo {address} {Singapore},\ \bibinfo
  {year} {2011})\BibitemShut {NoStop}%
\bibitem [{\citenamefont {Misner}(1969)}]{MisnerMixmaster}%
  \BibitemOpen
  \bibfield  {author} {\bibinfo {author} {\bibfnamefont {C.~W.}\ \bibnamefont
  {Misner}},\ }\href {https://doi.org/10.1103/PhysRevLett.22.1071} {\bibfield
  {journal} {\bibinfo  {journal} {Phys. Rev. Lett.}\ }\textbf {\bibinfo
  {volume} {22}},\ \bibinfo {pages} {1319} (\bibinfo {year}
  {1969})}\BibitemShut {NoStop}%
\bibitem [{\citenamefont {Kasner}(1921)}]{Kasner}%
  \BibitemOpen
  \bibfield  {author} {\bibinfo {author} {\bibfnamefont {E.}~\bibnamefont
  {Kasner}},\ }\href {https://doi.org/10.2307/2370192} {\bibfield  {journal}
  {\bibinfo  {journal} {Am. J. Math.}\ }\textbf {\bibinfo {volume} {43}},\
  \bibinfo {pages} {217} (\bibinfo {year} {1921})}\BibitemShut {NoStop}%
\bibitem [{\citenamefont {Belinskii}\ \emph {et~al.}(1982)\citenamefont
  {Belinskii}, \citenamefont {Khalatnikov},\ and\ \citenamefont
  {Lifshitz}}]{BKL}%
  \BibitemOpen
  \bibfield  {author} {\bibinfo {author} {\bibfnamefont {V.}~\bibnamefont
  {Belinskii}}, \bibinfo {author} {\bibfnamefont {I.}~\bibnamefont
  {Khalatnikov}},\ and\ \bibinfo {author} {\bibfnamefont {E.}~\bibnamefont
  {Lifshitz}},\ }\href {https://doi.org/10.1080/00018738200101428} {\bibfield
  {journal} {\bibinfo  {journal} {Adv. Phys.}\ }\textbf {\bibinfo {volume}
  {31}},\ \bibinfo {pages} {639} (\bibinfo {year} {1982})}\BibitemShut
  {NoStop}%
\bibitem [{\citenamefont {Taub}(1951)}]{Taub:1950ez}%
  \BibitemOpen
  \bibfield  {author} {\bibinfo {author} {\bibfnamefont {A.~H.}\ \bibnamefont
  {Taub}},\ }\href {https://doi.org/10.2307/1969567} {\bibfield  {journal}
  {\bibinfo  {journal} {Annals Math.}\ }\textbf {\bibinfo {volume} {53}},\
  \bibinfo {pages} {472} (\bibinfo {year} {1951})}\BibitemShut {NoStop}%
\bibitem [{\citenamefont {Segreto}\ and\ \citenamefont
  {Montani}(2024)}]{SebyGUP}%
  \BibitemOpen
  \bibfield  {author} {\bibinfo {author} {\bibfnamefont {S.}~\bibnamefont
  {Segreto}}\ and\ \bibinfo {author} {\bibfnamefont {G.}~\bibnamefont
  {Montani}},\ }\href {https://doi.org/10.1140/epjc/s10052-024-13145-2}
  {\bibfield  {journal} {\bibinfo  {journal} {Eur. Phys. J. C}\ }\textbf
  {\bibinfo {volume} {84}},\ \bibinfo {pages} {796} (\bibinfo {year} {2024})},\
  \Eprint {https://arxiv.org/abs/2401.17113} {arXiv:2401.17113 [gr-qc]}
  \BibitemShut {NoStop}%
\bibitem [{\citenamefont {Motaharfar}\ \emph {et~al.}(2024)\citenamefont
  {Motaharfar}, \citenamefont {Singh},\ and\ \citenamefont
  {Thareja}}]{noFriedmannLQC}%
  \BibitemOpen
  \bibfield  {author} {\bibinfo {author} {\bibfnamefont {M.}~\bibnamefont
  {Motaharfar}}, \bibinfo {author} {\bibfnamefont {P.}~\bibnamefont {Singh}},\
  and\ \bibinfo {author} {\bibfnamefont {E.}~\bibnamefont {Thareja}},\ }\href
  {https://doi.org/10.1103/PhysRevD.109.086013} {\bibfield  {journal} {\bibinfo
   {journal} {Phys. Rev. D}\ }\textbf {\bibinfo {volume} {109}},\ \bibinfo
  {pages} {086013} (\bibinfo {year} {2024})},\ \Eprint
  {https://arxiv.org/abs/2311.08465} {arXiv:2311.08465 [gr-qc]} \BibitemShut
  {NoStop}%
\bibitem [{\citenamefont {de~Cesare}\ and\ \citenamefont
  {Wilson-Ewing}(2019)}]{deCesare}%
  \BibitemOpen
  \bibfield  {author} {\bibinfo {author} {\bibfnamefont {M.}~\bibnamefont
  {de~Cesare}}\ and\ \bibinfo {author} {\bibfnamefont {E.}~\bibnamefont
  {Wilson-Ewing}},\ }\href {https://doi.org/10.1088/1475-7516/2019/12/039}
  {\bibfield  {journal} {\bibinfo  {journal} {JCAP}\ }\textbf {\bibinfo
  {volume} {12}}\bibfield  {number} {\bibinfo  {number} { (2019)},\ \bibinfo
  {pages} {039}},\ }\Eprint {https://arxiv.org/abs/1910.03616}
  {arXiv:1910.03616 [gr-qc]} \BibitemShut {NoStop}%
\bibitem [{\citenamefont {Rovelli}(1991)}]{Rovelli:1990ph}%
  \BibitemOpen
  \bibfield  {author} {\bibinfo {author} {\bibfnamefont {C.}~\bibnamefont
  {Rovelli}},\ }\href {https://doi.org/10.1088/0264-9381/8/2/011} {\bibfield
  {journal} {\bibinfo  {journal} {Class. Quant. Grav.}\ }\textbf {\bibinfo
  {volume} {8}},\ \bibinfo {pages} {297} (\bibinfo {year} {1991})}\BibitemShut
  {NoStop}%
\bibitem [{\citenamefont {Tambornino}(2012)}]{relationalobs}%
  \BibitemOpen
  \bibfield  {author} {\bibinfo {author} {\bibfnamefont {J.}~\bibnamefont
  {Tambornino}},\ }\href {https://doi.org/10.3842/SIGMA.2012.017} {\bibfield
  {journal} {\bibinfo  {journal} {SIGMA}\ }\textbf {\bibinfo {volume} {8}},\
  \bibinfo {pages} {017} (\bibinfo {year} {2012})},\ \Eprint
  {https://arxiv.org/abs/1109.0740} {arXiv:1109.0740 [gr-qc]} \BibitemShut
  {NoStop}%
\bibitem [{\citenamefont {Segreto}\ and\ \citenamefont
  {Montani}(2025)}]{Segreto:2024gcg}%
  \BibitemOpen
  \bibfield  {author} {\bibinfo {author} {\bibfnamefont {S.}~\bibnamefont
  {Segreto}}\ and\ \bibinfo {author} {\bibfnamefont {G.}~\bibnamefont
  {Montani}},\ }\href {https://doi.org/10.1088/1475-7516/2025/03/061}
  {\bibfield  {journal} {\bibinfo  {journal} {JCAP}\ }\textbf {\bibinfo
  {volume} {03}}\bibfield  {number} {\bibinfo  {number} { (2025)},\ \bibinfo
  {pages} {061}},\ }\Eprint {https://arxiv.org/abs/2407.20476}
  {arXiv:2407.20476 [gr-qc]} \BibitemShut {NoStop}%
\bibitem [{\citenamefont {Barca}\ and\ \citenamefont
  {Giovannetti}(2025)}]{Barca:2024goe}%
  \BibitemOpen
  \bibfield  {author} {\bibinfo {author} {\bibfnamefont {G.}~\bibnamefont
  {Barca}}\ and\ \bibinfo {author} {\bibfnamefont {E.}~\bibnamefont
  {Giovannetti}},\ }\href {https://doi.org/10.3390/universe11020063} {\bibfield
   {journal} {\bibinfo  {journal} {Universe}\ }\textbf {\bibinfo {volume}
  {11}},\ \bibinfo {pages} {63} (\bibinfo {year} {2025})},\ \Eprint
  {https://arxiv.org/abs/2412.20983} {arXiv:2412.20983 [gr-qc]} \BibitemShut
  {NoStop}%
\bibitem [{\citenamefont {Bergeron}\ \emph {et~al.}(2015)\citenamefont
  {Bergeron}, \citenamefont {Dapor}, \citenamefont {Gazeau},\ and\
  \citenamefont {Ma{\l}kiewicz}}]{Bergeron:2015jpa}%
  \BibitemOpen
  \bibfield  {author} {\bibinfo {author} {\bibfnamefont {H.}~\bibnamefont
  {Bergeron}}, \bibinfo {author} {\bibfnamefont {A.}~\bibnamefont {Dapor}},
  \bibinfo {author} {\bibfnamefont {J.~P.}\ \bibnamefont {Gazeau}},\ and\
  \bibinfo {author} {\bibfnamefont {P.}~\bibnamefont {Ma{\l}kiewicz}},\ }\href
  {https://doi.org/10.1103/PhysRevD.91.124002} {\bibfield  {journal} {\bibinfo
  {journal} {Phys. Rev. D}\ }\textbf {\bibinfo {volume} {91}},\ \bibinfo
  {pages} {124002} (\bibinfo {year} {2015})},\ \bibinfo {note} {[Addendum:
  Phys. Rev. D 91, 129905 (2015)]},\ \Eprint {https://arxiv.org/abs/1501.07718}
  {arXiv:1501.07718 [gr-qc]} \BibitemShut {NoStop}%
\bibitem [{\citenamefont {Ma{\l}kiewicz}\ \emph {et~al.}(2020)\citenamefont
  {Ma{\l}kiewicz}, \citenamefont {Peter},\ and\ \citenamefont
  {Vitenti}}]{Malkiewicz:2019azw}%
  \BibitemOpen
  \bibfield  {author} {\bibinfo {author} {\bibfnamefont {P.}~\bibnamefont
  {Ma{\l}kiewicz}}, \bibinfo {author} {\bibfnamefont {P.}~\bibnamefont
  {Peter}},\ and\ \bibinfo {author} {\bibfnamefont {S.~D.~P.}\ \bibnamefont
  {Vitenti}},\ }\href {https://doi.org/10.1103/PhysRevD.101.046012} {\bibfield
  {journal} {\bibinfo  {journal} {Phys. Rev. D}\ }\textbf {\bibinfo {volume}
  {101}},\ \bibinfo {pages} {046012} (\bibinfo {year} {2020})},\ \Eprint
  {https://arxiv.org/abs/1911.09892} {arXiv:1911.09892 [gr-qc]} \BibitemShut
  {NoStop}%
\bibitem [{\citenamefont {Ma{\l}kiewicz}\ \emph {et~al.}(2024)\citenamefont
  {Ma{\l}kiewicz}, \citenamefont {Ostrowski},\ and\ \citenamefont
  {Delgado~Gaspar}}]{MalkiewiczBianchiCMB}%
  \BibitemOpen
  \bibfield  {author} {\bibinfo {author} {\bibfnamefont {P.}~\bibnamefont
  {Ma{\l}kiewicz}}, \bibinfo {author} {\bibfnamefont {J.~J.}\ \bibnamefont
  {Ostrowski}},\ and\ \bibinfo {author} {\bibfnamefont {I.}~\bibnamefont
  {Delgado~Gaspar}},\ }\href {https://doi.org/10.1103/PhysRevD.109.083525}
  {\bibfield  {journal} {\bibinfo  {journal} {Phys. Rev. D}\ }\textbf {\bibinfo
  {volume} {109}},\ \bibinfo {pages} {083525} (\bibinfo {year}
  {2024})}\BibitemShut {NoStop}%
\bibitem [{\citenamefont {Agullo}\ \emph {et~al.}(2013)\citenamefont {Agullo},
  \citenamefont {Ashtekar},\ and\ \citenamefont {Nelson}}]{Agullo:2013ai}%
  \BibitemOpen
  \bibfield  {author} {\bibinfo {author} {\bibfnamefont {I.}~\bibnamefont
  {Agullo}}, \bibinfo {author} {\bibfnamefont {A.}~\bibnamefont {Ashtekar}},\
  and\ \bibinfo {author} {\bibfnamefont {W.}~\bibnamefont {Nelson}},\ }\href
  {https://doi.org/10.1088/0264-9381/30/8/085014} {\bibfield  {journal}
  {\bibinfo  {journal} {Class. Quant. Grav.}\ }\textbf {\bibinfo {volume}
  {30}},\ \bibinfo {pages} {085014} (\bibinfo {year} {2013})},\ \Eprint
  {https://arxiv.org/abs/1302.0254} {arXiv:1302.0254 [gr-qc]} \BibitemShut
  {NoStop}%
\bibitem [{arX()}]{arXivversion}%
  \BibitemOpen
  \href@noop {} {\bibinfo {title} {{ The supporting data for this article (in
  the form of a {\em Mathematica} notebook used to generate numerical solutions
  for Figure 4, created by Gabriele Barca in 2025) are openly available from
  the {\tt arXiv} version of the paper under
  \href{https://arxiv.org/abs/2507.01678}{{\tt
  https://arxiv.org/abs/2507.01678}}}.}}\BibitemShut {Stop}%
\end{thebibliography}%

\end{document}